\documentclass[showpacs,prl,onecolumn,aps,superscriptaddress,preprintnumbers,nofootinbib]{revtex4}
\usepackage[T1]{fontenc}
\usepackage[latin9]{inputenc}
\usepackage{amsmath,amssymb}
\usepackage{epsfig}
\usepackage{graphicx}
\usepackage{mathrsfs}
\usepackage{amsmath}
\usepackage{amsfonts}
\usepackage{epstopdf}
\usepackage{color}
\def\slashchar#1{\setbox0=\hbox{$#1$}     		
   \dimen0=\wd0                                 	
   \setbox1=\hbox{/} \dimen1=\wd1               	
   \ifdim\dimen0>\dimen1                        	
      \rlap{\hbox to \dimen0{\hfil/\hfil}}      	
      #1                                        	
   \else                                        	
      \rlap{\hbox to \dimen1{\hfil$#1$\hfil}}   	
      /                                         	
   \fi}

\newcommand{\beq}{\begin{equation}}
\newcommand{\eeq}{\end{equation}}
\newcommand{\bea}{\begin{eqnarray}}
\newcommand{\eea}{\end{eqnarray}}
\newcommand{\ba}{\begin{array}}
\newcommand{\ea}{\end{array}}

\def\eq#1{{Eq.~(\ref{#1})}}
\def\fig#1{{Fig.~\ref{#1}}}

\newcommand{\nn}{\nonumber}

\newcommand{\Lb}{\left(}
\newcommand{\Rb}{\right)}
\newcommand{\h}{\frac{1}{2}}

\newcommand{\pom}{I\!\!P}
\newcommand{\Y}{\tilde{Y}}

\newcommand{\intl}{\int\limits}
\begin{document}

\vskip1cm

\title{\bf Particle production in the toy world: multiplicity distribution and entropy}

\author{Eugene Levin}
\email{leving@tauex.tau.ac.il}
\affiliation{Department of Particle Physics, Tel Aviv University, Tel Aviv 69978, Israel}

\date{\today}


\begin{abstract}

      In this paper we found the multiplicity distribution of the produced dipoles in the final state  for dipole-dipole scattering in the zero dimension toy models. This distribution shows the great differences 
       from the distributions of  partons in the wave function of the projectile. However, in spite  of this difference the entropy of the produced dipoles turns out to be the same as the entropy of the dipoles in the wave function.  This fact is not surprising since  in the parton approach only dipoles in the hadron wave function which  can be produced at $t = +\infty$ and measured by the detectors.
We can also confirm the result of Kharzeev and Levin that this entropy is equal to $S_E = \ln\Lb xG(x)\Rb$, where we denote by $xG$ the mean multiplicity of the dipoles in  the deep inelastic scattering. The evolution equations for $\sigma_n$  are derived.

 \end{abstract}
\maketitle
\vspace{-0.5cm}
\tableofcontents

\section{Introduction}

Zero transverse dimension toy models can be viewed as a realization of the Pomeron calculus or more generally of 
Reggeon Field Theory (RFT). Over the years they have been intensively used to model high energy collisions in QCD. 
These models\cite{ACJ,AAJ,JEN,ABMC,CLR,CIAF,MUSA,nestor,RS,SHXI,KOLEV,BMMSX,BIT,LEPRI,utm,utmm,utmp} encode various fundamental features of QCD such as unitarity, but are much simpler than the latter and frequently 
solvable analytically.  Hence they provide a valuable playground to gain intuition about the dynamics of real QCD. 

In this paper, we continue  the exploration of the toy world following our early papers on the subject, Ref. \cite{utm,utmm,utmp}.
We consider dipole-dipole scattering and focus on particle production in the final state.  There are two motivations behind this work.
First, multiplicity distribution of produced particles is an important experimental observable measured in most of the experiments ranging 
from  DIS to heavy ion collisions\footnote{It is almost impossible to collect all references on the measurement of multiplicity.  We concentrated mostly on high energy $ p p$ collisions since in the paper we consider this particular process.}\cite{H1,H1MULT1,H1MULT2,ZEUSMULT,ATLAS1,ATLAS2,ATLAS3,ATLAS4,ATLAS5,KUTS,CMS1,CMS2,CMS3,CMS4,CMS5,TOTEM1,TOTEM2,ALICE1,ALICE2,ALICE3,ALICE4,ALICE5,ALICE6,LHCb,LHCb1,NA22,UA11,UA51,UA52,CDF1,CDF2,CDF3,RSH}. Particularly, measuring the multiplicity distribution of soft hadrons in heavy ion collisions
was instrumental to identify thermolization and hydrodynamization of produced QGP. 
A quantity tightly related to the  multiplicity distribution is the entropy of produced particles. It is believed to provide a valuable insight about particle production mechanism, particularly about the decoherence processes taking place throughout the scattering process. There are quite a few theoretical approaches to  entropy production 
in high energy collisions \cite{KUT,PES,KOLU1,PESE,KHLE,BAKH,BFV,HHXY,KOV1,GOLE1,GOLE2,KOV2,NEWA,LIZA,FPV,TKU,KOV3,KOV4,Hentschinski:2022rsa,DVVE,DVA1,DVA2}, 
which we will not review here. Experimental works on the subject have been reported too \cite{H1}.

As the second motivation behind this work, we were triggered by the recent ideas put forward in \cite{DVVE,DVA1,DVA2}. 
There, the $2\rightarrow n$ process has been identified with formation of a maximal entropy classical state, the {\it saturon}. 
Furthermore, this classical state was speculated to mimic a black hole formation through two graviton collision. 
One of the key assumptions of \cite{DVVE,DVA1,DVA2} is about the large $n$ behavior of the $2\rightarrow n$ cross section, $\sigma\Lb 2 \to n\Rb$.

Having a realistic  toy model of particle production at hand makes it possible to check the main assumption of these papers.

Our discussion starts (Section 2) from reviewing two toy models.  The first one is the BFKL cascade model  which mimics the Mueller's dipole model in QCD \cite{MUDI}. This model lacks both $s$ and $t$-channel unitarity \cite{KLremark2,utm} and hence has to be corrected. A more 
realistic  model is the Unitary Toy Model (UTM) first introduced in \cite{MUSA} and later explored in \cite{KLremark2,BIT,utm,utmm}.  These past studies have been mainly focused on 
formulation of RFT and computation of total $S$-matrix. 

Particle production in the toy world was originally addressed by Mueller and Salam in \cite{MUSA}. Using the AGK cutting rules \cite{AGK} 
applied to the BFKL cascade model, they derived a formula (denoted as MS formula below) for multiplicity distribution  
of produced particles (dipoles).
  The MS formula turns out to be frame dependent (violates the
 $t$-channel unitarity) and hence requires revision. Our prime goal in this paper is precisely to fill this gap.    
  We believe that all defects of the MS formula stem from the violation of t-channel unitarity in the BFKL cascade model.  Hence, we are going to discuss mainly the UTM  which is free from these shortcomings. While the total $S$-matrix is based on the knowledge of elastic amplitudes only,   particle production requires additional information 
about inelastic processes.   
  Our basic ideas are the same as in Ref.\cite{MUSA}: we are going to explore the fact that  the scattering amplitudes in all known zero dimension models can be calculated in the Pomeron calculus. First, we recall that it is proven in Refs.\cite{BFKL,LEHP} that the $s$-channel unitarity for the BFKL Pomeron has the form:
  \beq \label{I1}
2\, { \rm Im} \,G_{\pom}\Lb Y\Rb\,\equiv\,\,2\,  \,N\Lb Y\Rb\,\,=\,\,\sigma^{\mbox{\tiny BFKL}}_{in}(Y)
\eeq
 where $G_{\pom}$ is the Green's function for the BFKL Pomeron and  $\sigma^{\mbox{\tiny BFKL}}_{in}(Y)$ is the inelastic cross sections of  produced gluons with mean multiplicity $\bar{n} = \Delta\,Y$ , where $\Delta$ is the intercept of the BFKL Pomeron. We also know that produced gluons have the Poisson distribution  with this mean multiplicity (see Ref.\cite{LEHP} and appendix A).
 
\begin{figure}
 	\begin{center}
 	\leavevmode
 		\includegraphics[width=12cm]{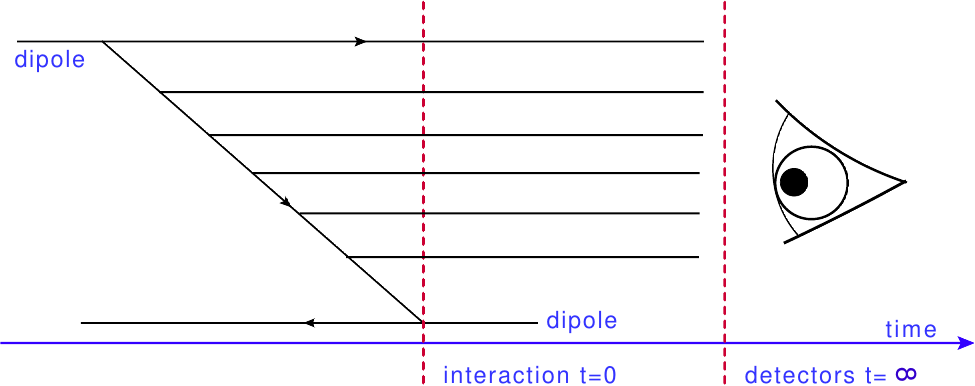}
				\end{center}
 	\caption{The sketch of dipole-dipole interaction in the parton approach. The lines denote pertons (dipoles),}
\label{pc}
 \end{figure}

 The second ingredient is the AGK cutting rules\cite{AGK}, which allow us to calculate the imaginary part of   the scattering amplitude, that determines the cross sections, through the powers of ${ \rm Im} \,G^{\mbox{\tiny BFKL}}\Lb Y\Rb$. Our master formula takes the form of convolution for the cross section of produced $n$  gluons:
 \beq \label{I2}
 \sigma_n\Lb Y \Rb\,\,=\,\,\sum_k\underbrace{ \sigma_k^{AGK}\Lb Y \Rb}_{ \propto\,\Lb{\rm Im} G_{\pom} \Rb^k}\underbrace{ \frac{\Lb \Delta\,k\,Y\Rb^n}{n!} e^{ - \Delta\,k\,Y}}_{\mbox{Poisson distribution}}\,\,\xrightarrow{Y \,\gg\,1} \,\, \sigma^{AGK}_{k = n/(\Delta\,Y)}\Lb  Y \Rb
\eeq 
   A mindful reader recognized that  our approach has been very successful in describing the  multiparticle production processes in the framework of soft Pomeron calculus ( see Ref.\cite{CANESCHI} ) and follows the main ideas of  Ref.\cite{MUSA}.  This approach has at least two advantages: (i) it can be easily generalized to the QCD case; and (ii) it replaces  in economic way a mess of parton interactions during  propagation of the parton cascade (see \fig{pc}) from the moment of interaction of wee parton with the target ($t = 0$ in \fig{pc})  till it reaches the detectors ($t = \infty$ in \fig{pc}).

  In section II  we mostly review  the results of Refs.\cite{utm,utmm,utmp}. The section III, IV and  V are  the main part of our paper in which discuss our approaches to  production of dipoles.  In section III  we compute the dipole multiplicity distributions in the final state based of the AGK cutting rules and the structure of the Pomeron exchange (see \eq{I1} and \eq{I2}).  This distributions show quite different behaviour in comparison with the multiplicity distribution in the wave function of the projectile.
  
  In section IV we generalize the approach of Refs.\cite{KOLEB,KOLE,CHMU,KLP,LEPRI} to the UTM cascade and show that the AGK cutting rules can be replaced by the evolution equations for $\sigma_n$.

In section V we find that the entropy of the produced gluons  turns out to be the same at the entropy of the dipoles in the wave  partonic function of the projectile.
 In conclusions we summarize our results.

 
 \section{ Scattering amplitudes in zero transverse dimensions models }
 
 Pomeron calculus in zero transverse dimensions
 models  QCD  for  dipoles of  fixed sizes. 
 Due to simplicity of the toy models of this type, they can be formulated as  Reggeon Field Theory (RFT) for  interacting bare Pomerons, and studied analytically. By bare Pomeron we mean a BFKL-like elastic scattering amplitude for dipole-dipole scattering  (below, we will also refer to "cut" Pomerons, which is a total inelastic cross section  in the same dipole-dipole process).
One hopes that the study of these models is a useful endeavor  assuming they retain
 some important features of real QCD.  
 
 Several models of this type have been recently considered in the literature. They all share the following simple probabilistic expression for 
 the total  $S$-matrix for scattering of $n$ dipoles of the projectile on $m$ dipoles of the target
  \beq \label{SMS}
 S(Y) \,\,=\,\,\sum_{n,m} e^{ - m \,n\,\gamma} \,P_n^P( Y_0)\,P_m^T( Y - Y_0)
 \eeq
 Here $\gamma \sim \alpha_s^2$ is the Born approximation to the dipole-dipole scattering amplitude at energies . 
 $P_n^P( Y_0) $ is the probability  to find $n$ dipoles  in the projectile boosted to rapidity $Y_0$ , and similarly for the  target. The probabilities are found by solving the evolution equation in rapidity, and it is in the form of this equation that the models differ from each other. The important consistency condition on the evolution is $t$-channel unitarity, which is essentially the requirement that  the $S$-matrix does not depend on  the reference frame, that is  the choice of $Y_0$ at a fixed $Y$\cite{MUSA,BIT,utm,utmm,utmp}.
 
 In this section we discuss the scattering amplitude  for the unitarity toy model (UTM).This model satisfies  both   the $t$ (see Refs.\cite{MUSA,BIT,utm}  and  $s$ (see Ref.\cite{utm} ) channel unitarity conditions.    
 Requiring $Y_0$ independence of the $S$-matrix in \eq{SMS} one is lead to the evolution equation for  $P^{\mbox{\tiny  UTM}}_n\Lb Y\Rb$ 
   \cite{MUSA,BIT}:
 \beq \label{MEQ}
\frac{d P^{\mbox{\tiny  UTM}}_n(Y)}{ d Y}\,=\,- \frac{\Delta}{1 - e^{-\gamma}} \Lb 1\,-\,e^{- \gamma n}\Rb P^{\mbox{\tiny  UTM}}_n(Y) \,\,+\,\,
\frac{\Delta}{1 - e^{-\gamma}} \Lb 1\,-\,e^{- \gamma(n - 1) }\Rb\,P^{\mbox{\tiny  UTM}}_{n-1}(Y)
\eeq  
 
 \eq{MEQ} can be rewritten as
 the general equation for the generating function of UTM:
 
 \beq\label{ZEQG}
\frac{\partial}{\partial \Y}Z(u, \Y)=-(1-u)\left(1-e^{-\gamma u\frac{\partial}{\partial u}}\right)Z(u  \Y)=\,( u - 1) \Bigg( Z(u,  \Y) - \,Z\Lb e^{-\gamma}\,u,\,  \Y \Rb\Bigg)
\eeq
where\cite{utmp}
\beq \label{ZEQG1}
Z\Lb u,  \Y \Rb\,\,=\,\,\,\sum^\infty_{n=0} C_n\Lb \gamma\Rb\,\,\Phi_n\Lb u,\gamma\Rb e^{ \Delta_n\,\Y}
\eeq 
$  \Phi_n\Lb u,\gamma \Rb $ are the eigenfunctions of the hamiltonian: ${\cal H} = -(1-u)\left(1-e^{-\gamma u\frac{\partial}{\partial u}}\right)$ with the eigenvalues $\Delta_n = \exp\Lb n\,\gamma\Rb \,-\,1$. $C_n\Lb \gamma\Rb = \exp\Lb \frac{n (n+1)}{2}\,\gamma\Rb$.  The scattering amplitude for dipole-dipole scattering takes the form\cite{utmp}:
  \beq \label{SA1}
S\Lb \Y\Rb\,\,=\,\,Z\Lb e^{-\gamma},\Y\Rb \,\,\,\,
\,=\,\,\,\sum^\infty_{n=0} C_n \Lb \gamma\Rb\,\Phi_n\Lb  e^{-\gamma},\gamma\Rb e^{ \Delta_n\,\Y}
\eeq
where we use \eq{ZEQG1}.  For small $\gamma$
$C_n\,\Phi_n\Lb 1 - \gamma,\gamma\Rb \,\,=\,\,\Lb - \gamma\Rb^n \,\,n!\,\,+\,\,O\Lb \gamma^{n+1}\Rb$, leading to
\beq \label{SA2}
S\Lb \Y\Rb\,\,=\,\,
\,\,\,\,\sum^\infty_{n=0} \Lb - \gamma\Rb^n \,\,n! \,\,e^{ \Delta_n\,\Y}\,\,\Bigg\{ 1\,\,\,+\,\,{\cal O}\Lb \gamma\Rb\Bigg\}
\eeq 
 
\eq{SA2} has  natural interpretation in the Pomeron calculus (see    appendix B and  \fig{dia}).

%
Using $n! = \int^\infty_0  d \tau e^{-\,\tau}\,\tau^n$ we can rewrite \eq{SA2} as follows:

\beq \label{SA31}
S\Lb \Y\Rb\,\,=\,\,
\,\,\,\,\intl^\infty_0 d\,\tau e^{- \tau} \sum^\infty_{n=0} \Lb - \gamma\,\tau\,\Rb^n \,\,e^{ \Delta_n\,\Y}
\eeq

\eq{SA2} and \eq{SA31} give the asymptotic series for the scattering amplitude. However, one can see that each term of this series has intercept which is large that $\gamma\,n (\Delta_n = e^{\gamma\,n} - 1 > \gamma\,n)$ and therefore, series of \eq{SA31} cannot be Borel summed. In Ref.\cite{utmp} it is suggested the following expansion for summing these series
\beq \label{EXP}
\exp\Lb \Delta_n\,\Y\Rb\,\,=\,\, e^{- \,\Y}\sum^\infty_{j=0} \frac{e^{\gamma \,n\,j}\,\Y^j}{j!}. 
\eeq
Plugging \eq{EXP} into \eq{SA31} we obtain
\beq \label{SA302}
S\Lb \Y\Rb\,\,=\,\,
\,\,\,\,\intl^\infty_0 d\,\tau e^{- \tau} \sum^\infty_{n=0} \Lb - \gamma\,\tau\,\Rb^n \,\, e^{- \,\Y}\sum^\infty_{j=0} \frac{e^{\gamma \,n\,j}\,\Y^j}{j!}\eeq
Both series are absolutely converged and changing the order of summation we have
\beq \label{SA303}
S\Lb \Y\Rb\,\,=\,\,
\,\,\,\,e^{- \,\Y}\sum^\infty_{j=0} \frac{\\Y^j}{j!}\intl^\infty_0 d\,\tau e^{- \tau} \sum^\infty_{n=0} \Lb - \gamma\,\tau\,\Rb^n e^{\gamma \,n\,j}\eeq
Summing over $n$ we obtain
\beq \label{SA304}
S\Lb \Y\Rb\,\,=\,\,
\,\,\,\,e^{- \,\Y}\sum^\infty_{j=0} \frac{\\Y^j}{j!}\intl^\infty_0 d\,\tau e^{- \tau} \frac{1}{ 1\,\,+\,\,\tau\,\gamma\,N_j}\,\,=\,\,\,\,e^{- \,\Y}\sum^\infty_{j=0} \frac{\Y^j}{j!}
\frac{1}{\gamma\,N_j} \exp\Lb\frac{1}{\gamma\,N_j}\Rb \,\Gamma\Lb 0,\frac{1}{\gamma N_j}\Rb\eeq
where $N_j = e^{ \gamma\,j}$.
 It is easy to see that the series in \eq{SA304} is absolutely converged and give the analytical function for the scattering amplitude. It is worthwhile mentioning that \eq{EXP} and \eq{SA304} give a way of summing the asymptotic series, which is quite different from the Borel summation, which is used to sum such series.

In Refs.\cite{utm,utmp} it was suggested a continuous approach to the UTM in which $P_n - P_{n-1} $ were replaced by $d P_n/d n$ and  a contribution of $d^2P_n/d n^2$ was neglected. In this approach
\eq{SA31} takes the form:
\bea \label{SA32}
S^{CA}\Lb \Y\Rb\,\,&=&\,\,
\,\,\,\,\intl^\infty_0 d\,\tau e^{- \tau} \sum^\infty_{n=0} \Lb - \gamma\,\tau\,\Rb^n \,\,e^{ \Delta \,n\,\Y}\,\,=\,\,
\,\,\,\,\intl^\infty_0 d\,\tau e^{- \tau} \frac{1}{ 1 \,\,+\,\,  \gamma\,\tau\, \,\,e^{ \Delta\,\Y}}\nn\\
\,\,&=&\,\,\,\frac{1}{\gamma\,N\Lb \Y\Rb} \exp\Lb\frac{1}{\gamma\,N\Lb \Y\Rb}\Rb \,\Gamma\Lb 0,\frac{1}{\gamma\,N\Lb \Y\Rb}\Rb
\eea
where $N\Lb \Y\Rb\,=\,\,e^{ \Delta\,\Y}$ and $\Gamma\Lb 0,z\Rb$ is incomplete Gamma function. It is worthwhile mentioning that in \eq{SA32} the Borel summation has been used.
 It is shown in Ref.\cite{utmp} that \eq{SA32}  can be reproduced directly from \eq{SMS} in the BFKL limit where in \eq{MEQ} we replace $e^{\gamma\,n} $ by $ 1 + \gamma\,n$.

 The scattering with the nuclear target has been discussed in Refs.
\cite{utmm,utmp}. From \eq{SA1}  this amplitude has the following form:
 
 \beq \label{SAA}
S^A\Lb \Y\Rb\,\,=\,\,Z\Lb e^{-\gamma\,A},\Y\Rb \,\,\,\,
\,=\,\,\,\sum^\infty_{n=0} C_n \Lb \gamma\Rb\,\Phi_n\Lb  e^{-\gamma\,A},\gamma\Rb e^{ \Delta_n\,\Y}
\eeq
In the kinematic region $\gamma \,A  \gtrsim \,1$ but $\gamma \,\ll\,1$ 
$ C_n \Lb \gamma\Rb\,\to\,1$ and $\Phi_n\Lb  e^{-\gamma\,A},\gamma\Rb\,\to\,\Lb - e^{\gamma\,A} \Rb^n$ leading to 
\beq \label{SAA1}
S^A\Lb \Y \Rb\,\,=\,\,\sum^\infty_{n=0}\Lb - e^{\gamma\,A\,} \Rb^n e^{ \Delta_n\,\Y}\,\,=\,\,
e^{- \,\Y} \sum^\infty_{j=0} \frac{\Y^j}{j!}\sum^\infty_{n=0}\Lb - e^{ \gamma \Lb j\,+\,A\Rb}\Rb^n\,\,=\,\,e^{- \,\Y} \sum^\infty_{j=0} \frac{\Y^j}{j!}\frac{1}{1\,+\,e^{\gamma\Lb j + A\Rb}}\eeq 
 in \eq{SAA} we assume that nucleus is the state of $A$ dipoles .

     \begin{figure}[ht]
    \centering
  \leavevmode
      \includegraphics[width=8cm]{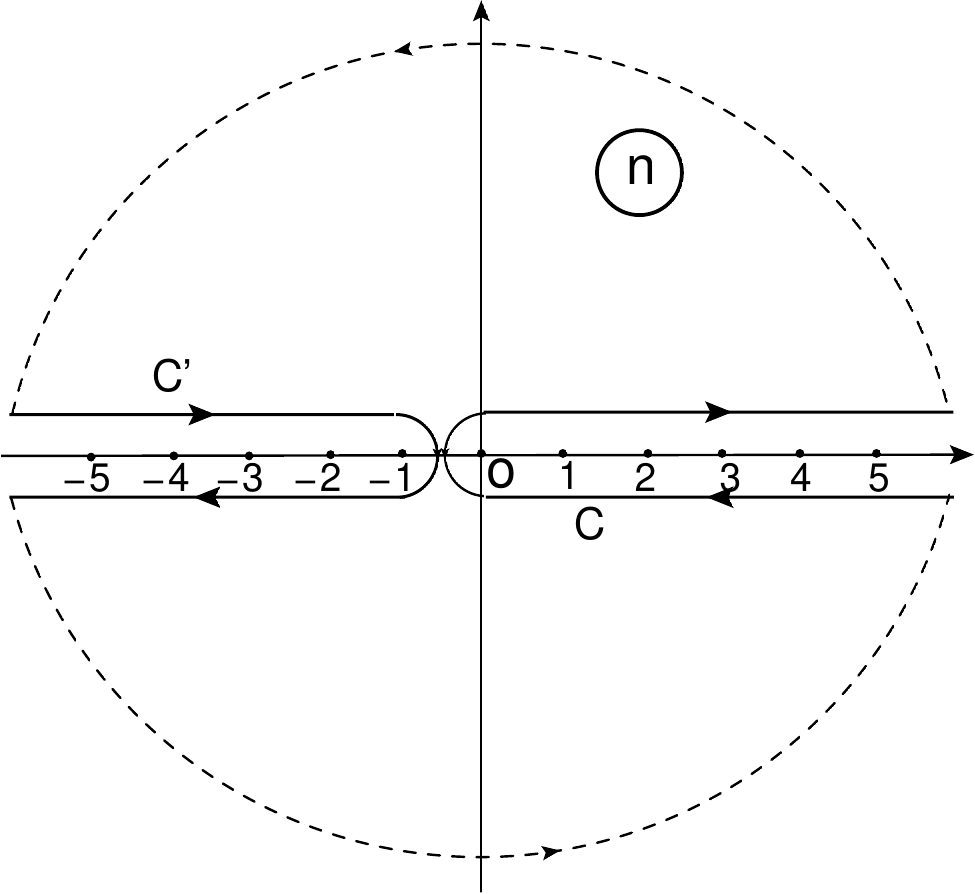} 
      \caption{ The contours of integrations in complex $n$ plane.}
\label{contr}
   \end{figure}
%
 
   \section{AGK cutting rules and  multiparticle production in the UTM}
   The scattering amplitude depends on the parton wave function of the fast projectile and 
   on $P_n$ (see \eq{SMS}) at  $\tau=0$ in \fig{pc}. Interaction of wee parton at $\tau=0$ destroys the coherence of partons in the wave function. Hence after this interaction the partons start to interact between themselves, these interactions  are difficult  to  tackle. In the introduction we have described our approach which is based on the AGK catting rules and on the proven fact that imaginary part of the Pomeron Green's function leads to the cross section of the produced gluons in QCD\footnote{ It should be mentioned that on QCD the partons at $\tau=0$  are colourless dipoles while at $\tau=\infty$ the gluons are produced.}.


  \begin{boldmath}
  \subsection{ Continuous approximation}
  \end{boldmath}
For arbitrary $Y$ we have to use the scattering amplitude in a general form of \eq{SA1} and \eq{SA304}, However, in the continuous approximation
we can restrict ourselves by \eq{SA32} which is much simpler. It has 
 has the following simple form
 \beq \label{SAG}
S\Lb Y\Rb\,\,=\,\,\sum_{k=0}^\infty (- 1)^{k} C_k\Lb \gamma\Rb [N\Lb Y\Rb]^k\,\,; ~~~~~~~ C_0=1
 \eeq 
with $C_k =  \,\gamma^k\,k!$ . $N\Lb Y\Rb$ denotes the imaginary part of the Pomeron Green's function (see \eq{I1}) and it equals to $\exp\Lb \Delta_1 \Y\Rb$.  In the toy world, as well as in QCD \cite{MUDI}  the imaginary part of the Pomeron Green's function coincides with  the multiplicity of dipoles in the BFKL cascade.

 Each term in \eq{SAG}  is a contribution  of $k$ BFKL Pomerons.
The Pomeron Green's  function $G_{\pom}(Y)$ depend on rapidity but not on $\gamma$, while the dependence on $\gamma$ comes through the coefficients $C_k$.
The latter have interpretation of multi-Pomeron residues (impact factors). The expansion (\eq{SAG}) is not general, because not every function $S(Y)$ is expandable in powers of $N( Y)  = {\rm Im} G_{\pom}(Y)$. A compact representation of  \eq{SAG} in the UTM has the form (see \eq{SA32} 
\beq \label{SMS5}
S^{CA}  \,=\,\intl^\infty_0 d t\, e^{-t} \frac{1}{1\,+\,t\,\gamma\, N\Lb Y\Rb}
\eeq 
 The Pomeron plays two fold role in the scattering amplitude. First, it gives the scattering amplitude of two dipoles at high energy. The interaction of the Pomerons with the colliding dipoles and between them provides the shadowing in the amplitude which has been taken into account in \eq{SMS5}. Second, the imaginary part of the Pomeron exchange due to the unitarity constraints of \eq{I1} give the cross section of produced gluons. The AGK cutting rules allows us to calculate the cross section proportional to $\Lb {\rm Im}G_{\pom}(Y) \Rb^n = \Lb\sigma^{\mbox{\tiny BFKL}}_{in}(Y)\Rb^n$ \footnote{ This contribution in  the widely accepted slang is called the contribution of $k$ cut Pomerons.} out of  the contribution to the total cross sections of $k$ exchanged Pomerons ($\propto G^k_{\pom} \Lb Y \Rb $). The AGK rules give:
   \begin{subequations} 
    \bea \label{AGK}
n\,\geq\,1:~~\sigma^k_n\Lb Y\Rb& =&C_k(- 2 )^{k-n}\frac{k!}{(k - n)!\,n!}\,\Lb \sigma^{\mbox{\tiny BFKL}}_{in}(Y)\Rb^n
\, \label{AGKK}\\
n\,=\,0:~~\sigma^k_0\Lb Y\Rb&=& \Bigg(2\,\,-\,\,2^k\Bigg)C_k \Big(- N\Lb Y\Rb\Big)^k ;\label{AGK0}
\eea
 \end{subequations} 
Here $n=0$ is the diffractive dissociation.

The total "production" of $n$ cut Pomerons is  the sum over all exchanges ($k\ge n$):
\bea\label {sn}
\sigma_n^{AGK} =\sum_{k=n} \sigma^k_n &=& 
\sum^\infty_{k=n}  C_k(- 2 )^{k-n}\frac{k!}{(k - n)!\,n!}\,\Lb  \sigma^{\mbox{\tiny BFKL}}_{in}(Y)\Rb^n \nn \\
&=&\sum^\infty_{k=n}  (-1)^{ k-n}C_k( 2\,N\Lb Y\Rb )^{k}\frac{k!}{(k - n)!\,n!}
\eea
where \eq{I1} was used to get the last expression.

While for the UTM model, the coefficients $C_k$ are known explicitly,  in general they  could  also be conveniently expressed as follows:
\beq
C_k= {(-1)^k\over k!} \,{d^k\over dN^k} \,S(Y)_{|_{N=0}}
\eeq
Substituting $C_k$ into \eq{sn},
\beq \label{AGKN}
 \sigma_n^{AGK} \Lb Y \Rb
 = \int^\infty_0 d t \,e^{-t} \frac{ \Lb 2\,t\,\gamma\,N \Lb Y\Rb\Rb^n }{\Big(1\,\, +\,\,2\,t\,\gamma\, N\,\Lb Y\Rb\Big)^{n + 1}}\,\,=\,\, \frac{1}{2\,\gamma\,N(Y)}k!\,U\left(k+1,1,\frac{1}{2 \gamma  N(Y)}\right) 
\eeq
 U(a,b,z) is the  Tricomi confluent hypergeometric function (see Ref.\cite{AS}, formula {\bf 13.1.3}).  
    We note two interesting limits of \eq{AGKN}:
   \bea \label{AGKN1}
\sigma_k(Y<Y_{max})\,\,=\,\,\frac{1}{2\,\gamma\,N}\,k!\, U\left(k+1,1,\frac{1}{2\,\gamma\,N}\right) \,\,\rightarrow\,\,
 \left\{\begin{array}{l}\,\,\,\frac{1}{\gamma\,N} \,\displaystyle{K_0\Bigg( 2 \sqrt{ \frac{k+1}{2\gamma\,N}}\Bigg)}\,\,\,\,\,\,\,\,\,\,\mbox{for}\,\,\,2 \gamma\, N \,k\,>\,1;\\ \\
\,\,\displaystyle{k! \Lb 2\,\gamma\, N\Rb^{k}} \,\,\,\,\,\mbox{for}\,\,\,2 \gamma\, N \,k\,<\,1;\\  \end{array}
\right.
 \eea 
 
   Coming back to \eq{AGKN1}  one can see that  we  indeed   have $ \sigma\Lb 2 \to n\Rb$, which behaves as $ \sigma\Lb 2 \to n\Rb=C_n n! \alpha^n$\cite{DVA1,DVA2} where $\alpha$ is the four boson coupling constant is small in the t'Hooft limit: $ \alpha \,\to\,0, \lambda_t = \alpha\,N_c  = {\rm finite}$.    
  However, $\alpha $ in \eq{AGKN1} is replaced by the BFKL contribution:  $   2\,\gamma\,N\Lb Y\Rb$ . Note that this contribution gives the scattering amplitude of two dipoles (four dipoles interaction) but depends on energy increasing with it. Due to this growth, at large energies 
  $\sigma_n \sim n! \Lb 2\,\gamma\, N\Rb^{n} $ does not contribute (see \eq{AGKN1})  and the unitarity is saturated by the states with multiplicities $n \geq N $.  None of these states look classical.

 Summing over $n$ (the sum starts from $n=1$) we obtain the total inelastic cross section
\beq
\sigma_{in}\,=\,\sum_{n=1}^\infty  \sigma_n^{AGK}\,=\, 1 - S(Y)|_{N\rightarrow 2N}  
\eeq


 
 In the BFKL formalism ( see appendix A), the inelastic cross section for production of 
 $n$ - partons ($ \sigma^{\mbox{\tiny BFKL}}_{n}(Y)$) from a single Pomeron obeys the evolution equation
 \beq \label{MDTM1}
 \frac{d \sigma^{\mbox{\tiny BFKL}}_{n}(Y) }{d\,Y}\,\,=\,\,\Delta\,\sigma^{\mbox{\tiny BFKL}}_{n-1}(Y) 
 \eeq
The solution takes the form:
 \beq \label{MDTM2}
  \sigma^{\mbox{\tiny BFKL}}_{n}(Y)\,\,=\,2\,\frac{\Lb\Delta\,Y\Rb^n}{n!}\, ;~~~~~~~~~~~~~\sigma_{in}^{BFKL}\,=\,\sum_{n=0}^\infty \sigma^{\mbox{\tiny BFKL}}_{n}(Y)\,=2\,\exp\Lb \Delta\,Y\Rb\,=\,2\,N
 \eeq 
 consistently with the unitarity condition of \eq{I1}. Here the factor 2 comes from the unitarity condition for $n=0$.
The probability ${\cal P}^{\pom}_n\Lb Y\Rb$ to find $n$ partons in the final state  is thus
  \beq \label{MDTM3}
{\cal P}^{\pom}_n\Lb Y\Rb\,=\,\,\frac{ \sigma^{\mbox{\tiny BFKL}}_{n}(Y)}{\sum_{n=0}^\infty \sigma^{\mbox{\tiny BFKL}}_{n}(Y)} \,\,=\,\,\frac{\Lb\Delta\,Y\Rb^n}{n!}e^{ - \Delta\,Y}
 \eeq  
which  is a Poisson distribution with the average number of partons  $\bar{n}_1 = \Delta\,Y=\ln N$.  This is the result for a single cut Pomeron.
For $k$ cut Pomerons, the distribution is  Poisson again, but now with the average number $\bar{n}_k =k\,\bar{n}_1 = k\, \Delta\,Y =\ln N^k $. This corresponds 
to probabilities ${\cal P}^{\mbox\tiny \pom}_n\Lb k\,Y\Rb$.
 
 To find the multiplicity distribution of particles in the final states  one needs to convolute  $\sigma_k \Lb Y \Rb$ from (\ref{AGKN}) with
 the distribution of particles inside $k$ cut Pomerons \eq{MDTM3}:
  \beq \label{AGKN2} 
\sigma^{f.s.}_n\Lb Y\Rb \,\,=\,\,\sum^{\infty}_{k=1}\sigma_{k}^{AGK}(Y) \,\,{\cal P}^{\pom}_n \Lb k\,Y\Rb  
 \eeq
 In \eq{AGKN2} f.s. stands for the final state.

 ~
 
    \begin{boldmath}
  \subsection{ Arbitrary $Y$}
  \end{boldmath}
For general approach we have to use the scattering amplitude in a general form of \eq{SA1} and \eq{SA304}.

The AGK cutting rules \cite{AGK} allows us to calculate the contributions of $n$-cut Pomerons if we know $F_k$: the contribution of the exchange of $k$-Pomerons to  the cross section. They take the form:
   \begin{subequations} 
    \bea \label{AGKK}
n\,\geq\,1:\sigma^k_n\Lb Y\Rb&=& (-1)^{k-n}\frac{k!}{(n - k)!\,n!}\,2^{k}\, F_k(\gamma, Y)\label{AGKK}\\
n\,=\,0:\sigma^k_0\Lb Y\Rb&=&\Lb -1\Rb^k \Bigg(2^k\,\,-\,\,2\Bigg) F_k(\gamma, Y);\label{AGK0}\\
\sigma_{tot}&=&\,\,2 \sum_{k=1}^\infty (-1)^{k+1} \,F_k(\gamma, Y);\label{XS}\,
\eea
 \end{subequations} 
where $\sigma_{tot}$ is the total cross section and $\sigma_0$ denotes the cross section with the multiplicity of produced dipoles  which is much less than $\Delta\,Y$. In other words, it is the cross section of the diffraction production.  From \eq{SA31} one can see that
\beq \label{XSK1}
F_k(\gamma,Y)\,\,=\,\,\intl^{\infty}_0 d \tau\,\, e^{-\tau} \Lb \tau\,\gamma\Rb^k \,e^{\Delta_k\,\Y} 
\eeq
Using \eq{EXP} we can rewrite \eq{XSK1} in the following form
\beq \label{XSK2}
F_k(\gamma,Y)\,\,=\,\,\,e^{- \,\Y}\sum^\infty_{j=0} \frac{e^{\gamma \,k\,j}\,\Y^j}{j!}\intl^{\infty}_0 d \tau\,\, e^{-\tau} \Lb \tau\,\gamma\Rb^k  
\eeq

Bearing in mind \eq{AGKK} we can introduced
\beq \label{XSK}
\sigma^{\rm AGK}_n\,\,=\,\,\sum_{k=n}^\infty(-1)^{k - n} \frac{k!}{(k - n)!\,n!}\,2^{k}\, F_k(\gamma,Y);~~~~~  \sigma^{\rm AGK}_{in} \,=\,\sum^\infty_{n=1} \sigma^{\rm AGK}_n 
\eeq  
which is the cross section of production of $n$-cut Pomerons in our process and $\sigma^{\rm AGK}_{in} $ is the inelastic cross section.

Plugging \eq{XSK1} into \eq{XSK} and summing over $k$ we obtain:
\beq \label{XSK2}
\sigma^{\rm AGK}_n\,\,=\,\,e^{- \,\Y}\sum^\infty_{j=0} \frac{\\Y^j}{j!}\,\intl^{\infty}_0 d \tau\,\, e^{-\tau}\frac{\Lb2\,\tau\,\gamma\,N_j\Rb^n}{\Lb1 \,\,+\,\,2\,\tau\gamma\,N_j\Rb^{n+1}}\,\,=\,\,\,e^{- \,\Y}\sum^\infty_{j=0} \frac{\Y^j}{j!}\,\,n! \frac{1}{\gamma\,N_j} \,U\Lb n + 1, 1,\frac{1}{\gamma\,N_j}\Rb
\eeq

 Using \eq{AGKN1} for $U\Lb n + 1, 1,\frac{1}{\gamma\,N_j}\Rb$ we can take the integral over $j$ in \eq{XSK3} using the method of steepest descent.  The equation for the saddle point value of j $j = j_{SP}$ takes the form:
 \beq \label{XSK3}
\ln \Y\,\,-\,\,\ln j_{SP} \,\,+\,\,\gamma\sqrt{\frac{k+1}{2\,\gamma\,N_{j_{SP}}}}\,\,=\,\,0
 \eeq
 The solution to \eq{XSK3} at large $\Y$ has the form:
 \beq \label{XSK31}
 j_{SP} \,\,=\,\,\Y\exp\Lb \gamma\sqrt{\frac{k+1}{2\,\gamma\,N\Lb \Y\Rb}}\Rb~~~~\mbox{with} ~~~~~N\Lb \Y\Rb= e^{ \gamma\,\Y}
 \eeq
 
Plugging \eq{XSK31} into \eq{XSK2} we obtain:
\beq \label{XSK4}
\sigma^{\rm AGK}_n\,\,=\,\,\frac{1}{\gamma\,N\Lb \Y\Rb}\exp\Lb - 2\,\, \sqrt{\frac{k \,+\,1}{2\,\gamma \,N\Lb \Y\Rb}} \,\,\,-\,\,\gamma^2\,\Y\,\frac{k+1}{2\,\gamma\,N\Lb\Y\Rb}\Rb
\eeq
In \eq{XSK4} we use the asymptotic expression for $K_0\Lb  2 \sqrt{\frac{k \,+\,1}{2\,\gamma \,N\Lb \Y\Rb}}\Rb $ for simplicity.     
    
     \begin{figure}[ht]
    \centering
  \leavevmode
  \begin{tabular}{c c }
  \includegraphics[width=8cm]{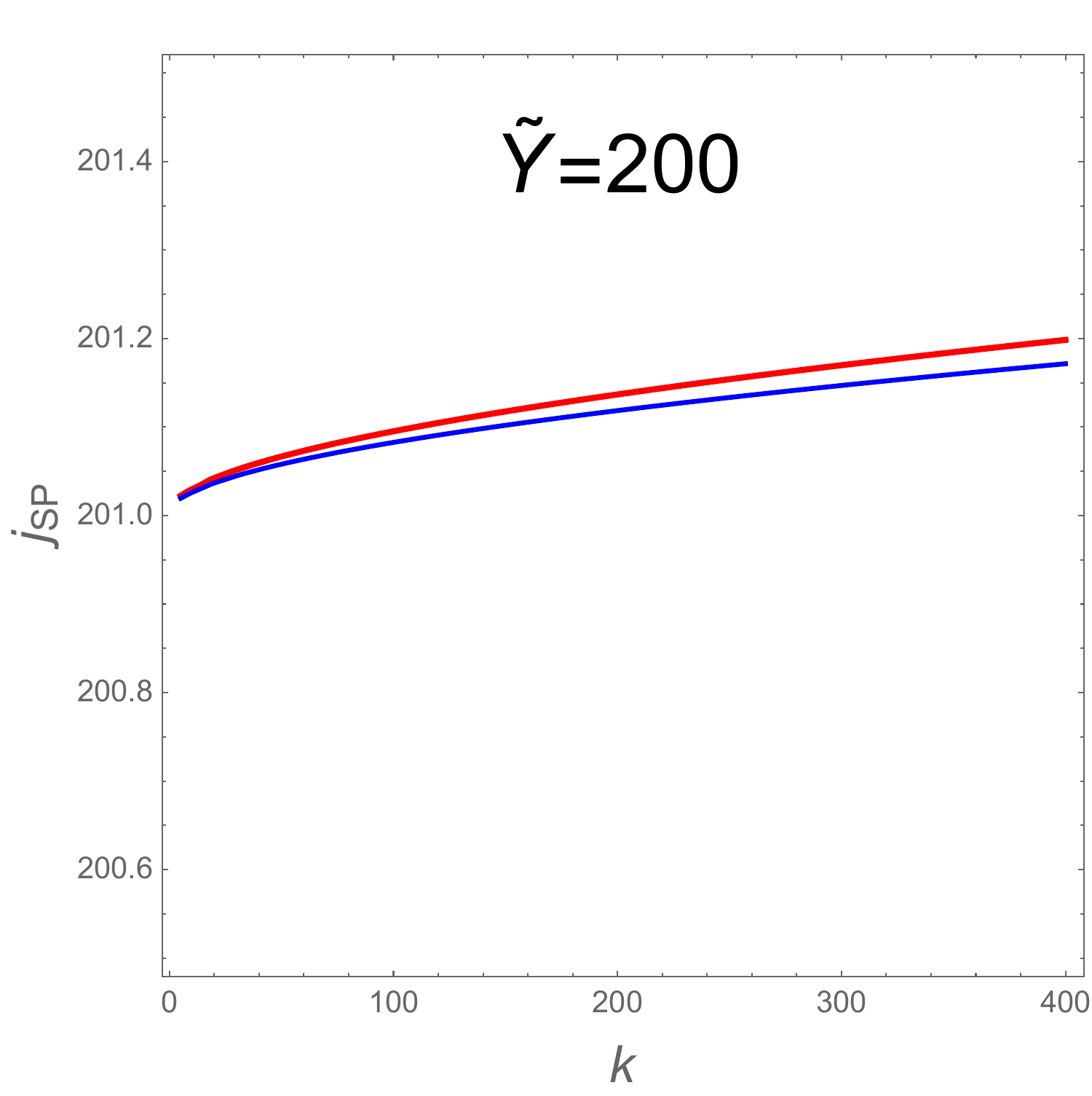}&  \includegraphics[width=7.8cm]{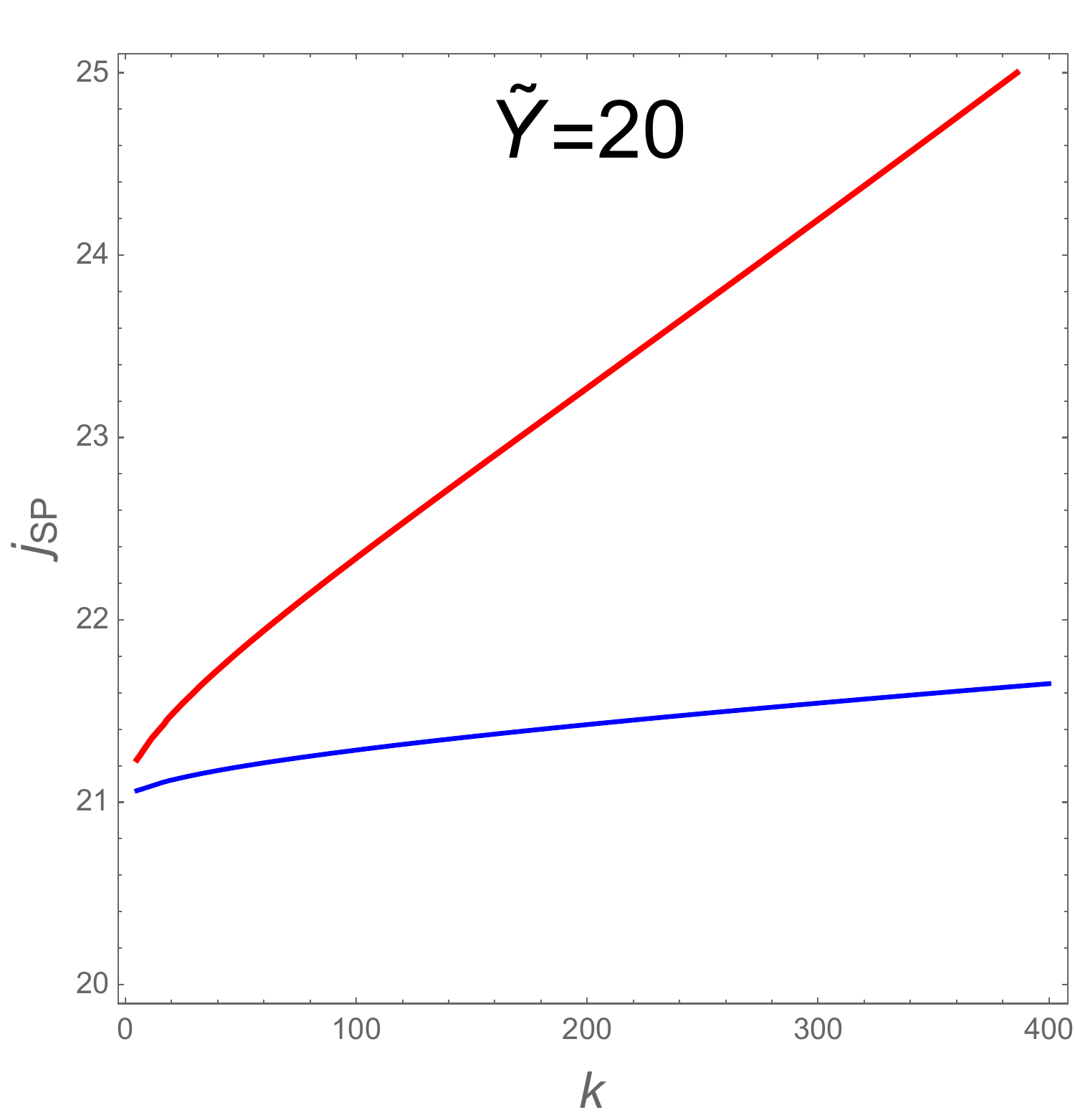}   \\
\end{tabular}
     \caption{ $j_{SP}$ of \eq{XSK3}  versus $n$ at different $Y$ (red curves). The blue curves show the saddle point value of $j_{SP}$ given by \eq{XSK31}. $\Delta=0.2$ and $\gamma = 0.025$. }
\label{utmjsp}
   \end{figure}

  \fig{utmjsp} illustrates the  difference between the exact solution to \eq{XSK3} and the approximate solution at large $\Y$.
  
  ~

      \begin{boldmath}
  \subsection{ Interaction with nuclei}
  \end{boldmath}
The scattering matrix for interaction with a nucleus target has been written in \eq{SAA1}.  Applying the  AGK cutting rules we obtain for $\sigma^{AGK, A}_n$ the following expression:
 \bea \label{XSK6}
 \sigma^{AGK, A}_n\,\,&=&\,\,e^{- \,\Y}\sum^\infty_{j=0} \frac{\Y^j}{j!}\,\,\frac{\Lb 2\,\gamma e^{\gamma \Lb j + A\Rb}\Rb^n}{\Lb1\,\,+\,\, 2\,\gamma e^{\gamma \Lb j + A\Rb}\Rb^{n+1}} \nn\\ & \,\xrightarrow{\gamma\,e^{\gamma \Lb j + A\Rb} \gg 1} & e^{- \,\Y}\sum^\infty_{j=0} \frac{\Y^j}{j!}\,\,\frac{1}{2\, \gamma\,e^{\gamma \Lb j + A\Rb}}\exp\Lb - \frac{n-1}{2\,\gamma\,e^{\gamma \Lb j + A\Rb }}\Rb
 \eea 
 \eq{XSK6} differs from the BFKL multiplicity distribution due to sum over $j$ which stems from the enhanced diagrams. Plugging in this equation $ j = \Y$ we obtain that 
 \beq \label{XSK61}
  \sigma^{AGK, A}_n =  \frac{1}{2\, \gamma\,e^{\gamma \Lb j\Y+ A\Rb}}\exp\Lb - \frac{n-1}{2\,\gamma\,e^{\gamma \Lb \Y + A\Rb }}\Rb 
  \eeq 
  which is the BFKL distribution with the mean multiplicity in $e^{\gamma \,A}$ times larger that in the BFKL cascade  for one dipole.

~

~

\begin{figure}[ht]
    \centering
  \leavevmode
      \includegraphics[width=10cm]{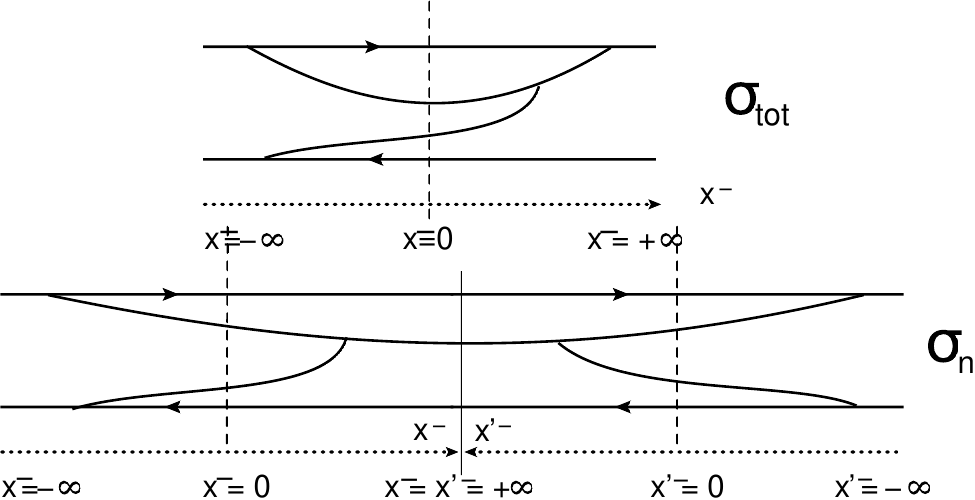}  
      \caption{ A comparison of calculations of  $\sigma_{tot}$ and $\sigma_n$ in parton approach. At $t=0$ the parton cascade interacts with the target. At $t = + \infty$  the produced partons are measured by the detectors.}
\label{timestr}
   \end{figure}

    \begin{boldmath}
    \section{Evolution of $\sigma_n$ in  parton approach}   
     \end{boldmath}
  It has been shown in Refs.\cite{KOLE,KOLEB} (see also Refs.\cite{LEPRI,KLP}) that instead of using the AGK cutting rules, we can write the evolution equation in  parton  approach.  This equation is based on two principle ingredients. The first one is that     only partons in the parton wave function of the fast hadron at $t \to - \infty$ in \fig{timestr} can be produced at    $t \to + \infty $ and are measured by our detectors. This point was proven in Refs.\cite{KOLE,KOLEB} and the proof is used the results of Ref.\cite{CHMU}. These papers  show that neither production of the new partons nor annihilation of the partons in the hadronic wave function contribute to  $\sigma_n$, including single diffraction. The proof is general and does not depend on the specific interaction between partons. Actually this feature allows us to apply parton approach for discussing the multiplicity distributions of produced partons. It has been demonstrated\cite{KOLE,LEPRI,KLP}  that the AGK cutting rules provides these features in the framework of the parton approach.
%

~

~
  \subsection{ BFKL cascade}

\begin{figure}[ht]
    \centering
  \leavevmode
      \includegraphics[width=18cm]{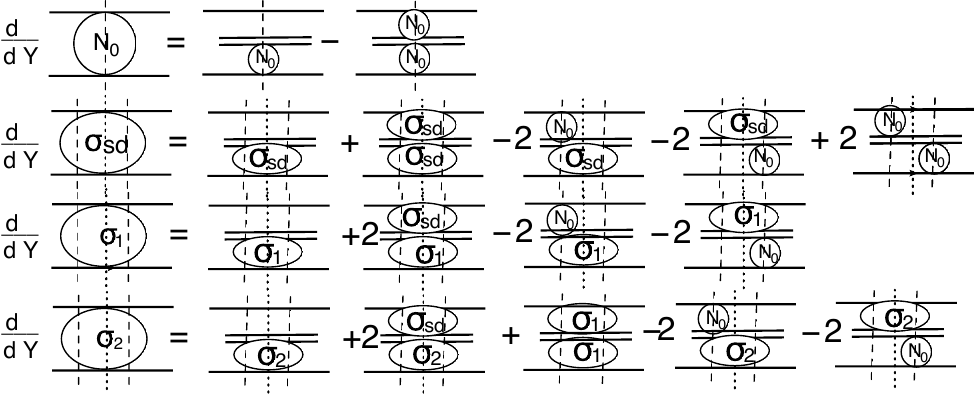}  
      \caption{ The graphic form of \eq{EVEQ1} - \eq{EVEQ4}. The vertical lines notations are  the same as in \fig{timestr}.  The Pomeron intercept is absorbed in the definition of $Y$.}
\label{eveq}
   \end{figure}

 The second principle feature has been proven for the BFKL parton cascade\cite{KOLE,KOLEB,LEPRI,KLP}  we can write the following evolution equations (see \fig{eveq}) which we present  for simplicity  for  the toy models\footnote{The proof for QCD is given in Refs.\cite{KOLE,KLP}.}:
  \begin{subequations} 
    \bea  
 \frac{ d N_0\Lb Y\Rb  }{d\,Y}& =&\Delta \Lb N_0\Lb Y\Rb \,\,-\,\,N^2_0\Lb Y\Rb\Rb; \label{EVEQ1}  \\
 \frac{d \sigma_{sd}(Y)}{\partial Y}\,&=&\,
\,\,\Delta \Lb \sigma_{sd}(Y)\,+\,\sigma^2_{sd}(Y)
\,-\,4\,\sigma_{sd}(Y) N_0(Y)
\,+\, 2
N^2_0(Y)\Rb; \label{EVEQ2} \\
 \frac{d \sigma_{1}(Y)}{d\,Y}\,&=&\,\Delta \Lb\sigma_1\Lb Y\Rb\,\,-\,\,2\,\sigma_1\Lb Y\Rb \Lb 2\,N_0\Lb Y\Rb \,-\,\sigma_{sd}\Lb Y\Rb\Rb\Rb; \label{EVEQ3} \\
 \frac{d \sigma_{2}(Y)}{d\, Y}\,&=&\,\Delta \Lb\sigma_2\Lb Y\Rb\,\,+\,\,2\,\sigma_2\Lb Y\Rb \sigma_{sd} \,+\,\sigma_1^2\Lb Y\Rb  \,-\,4\,N_0\Lb Y\Rb \sigma_2\Lb Y\Rb\Rb; \label{EVEQ4}\\
  \frac{d \sigma_{n}(Y)}{d\, Y}\,&=&\,\Delta \Lb\sigma_n\Lb Y\Rb\,\,+\,\,2\,\sigma_n\Lb Y\Rb \sigma_{sd} \,+\,\sum_{k=1}^n \sigma_{n-k}\Lb Y\Rb\sigma_k\Lb Y\Rb  \,-\,4\,N_0\Lb Y\Rb \sigma_n\Lb Y\Rb\Rb; \label{EVEQ5}  
    \eea    
  \end{subequations}  
 where  $ \Delta$ is the intercept of the BFKL Pomeron,  $N_0$ is the imaginary part of the elastic amplitude, $\sigma_{sd}$ is the cross section of the single diffraction, and $\sigma_n$ is the cross section of production of $n \Delta \,Y$ number of partons.

   The graphic forms of these equations  are shown in \fig{eveq}. It turns out that all numerical factors in this figure reproduce the AGK cutting rules.  In this figure we consider that partons in our approach are dipoles of QCD with the fixed sizes.  Every equation in \fig{eveq} is written for the cross section of interaction of one dipole  shown by two lines with the target which is presented in the figure by vertical dotted line.  This dipole decays in two dipoles which interact with the target. In the figure all possible interactions are shown with trivial factors that take into account the number of possible interactions and the shadowing due to interaction of elastic amplitude ($N_0$) with the cross sections in the equations: this shadowing gives the sign minus. 
   

       \subsection{ UTM cascade}

   Our main goal to study what happens with \eq{EVEQ1}-\eq{EVEQ5} in the UTM cascade.


       \subsubsection{ Elastic amplitude}

 Let us start from \eq{EVEQ1} for the elastic amplitude.  We can find $\frac{d\,S\Lb Y \Rb}{d \,Y}$ using \eq{SMS} for $Y_0=0$ since in this model S-matrix does not depend on $Y_0$. In doing so, we obtain the following equation for the scattering of two dipoles:
 \bea \label{EVEQST1}
 \frac{d \,S\Lb Y\Rb}{d\,Y}\,\,&=&\,\,\sum_{n=1}e^{ - \gamma\,n}\frac{d P^{\mbox{\tiny UTM}}_n\Lb Y\Rb}{d\,Y} \,\,=\,\,\frac{\Delta}{\gamma}\Lb e^{\gamma} \,-\,1\Rb \sum_{n=1}\Bigg(- e^{ - \gamma\,n} \,P^{\mbox{\tiny UTM}}_n\Lb Y\Rb \,\,+\,\,e^{ -2 \gamma\,n} P^{\mbox{\tiny UTM}}_n\Lb Y\Rb\,\Bigg)\nn\\
 &=&\,\frac{\Delta}{\gamma}\Lb e^{\gamma} \,-\,1\Rb \Bigg( S_{d + 2 d} \Lb Y\Rb \,\,-\,\,S_{d+d} \Lb Y\Rb\Bigg)
 \eea
 where $S_{d+d} $ is the scattering matrix of two dipoles while $S_{d+ 2 d}$ is the scattering matrix for interaction of one dipole with two dipoles.   It is instructive to derive this formula from \eq{SA2} for the scattering matrix. Indeed, 
  \bea \label{EVEQST2}
 \frac{d \,S\Lb Y\Rb}{d\,Y}\,\,&=& \sum^\infty_{n=0}\,\Delta_n C_n \Lb \gamma\Rb\,\Phi_n\Lb  e^{-\gamma},\gamma\Rb e^{ \Delta_n\,\Y} =\frac{\Delta}{\gamma}\sum^\infty_{n=0}\,\Lb e^{\gamma \,n}\,-\,1\Rb\,C_n \Lb \gamma\Rb\,\Phi_n\Lb  e^{-\gamma},\gamma\Rb e^{ \Delta_n\,\Y} \eea
 In Ref.\cite{utmp} we found that $C^{(2)}_n$ for the scattering amplitude with two dipoles are equal to
   \beq \label{EVEQST3} 
   C^{(2)}_n   \,\,=\,\,\frac{e^{ \gamma (n+1)}\,-\,1}{e^\gamma\,-\,1} \,\,\underbrace{C^{(1)}_n}_{\equiv C_n}\,\,\,\xrightarrow{\gamma \ll 1}\,\,\,\Lb n+1\Rb \,C^{(1)}_n   \eeq

   Plugging \eq{EVEQST3} into \eq{EVEQST2}  we obtain \eq{EVEQST1}.
   Comparing \eq{EVEQST1} with \eq{EVEQ1} one can see that in the BFKL cascade 
   \beq \label{EVEQST4}
   S_{d + 2 d} \Lb Y\Rb \,\,=\,\,S^2_{d + d} \Lb Y\Rb
   \eeq
   Using  \eq{EXP}, \eq{SA2} and \eq{EVEQST3} we can calculate $S_{d + 2d}$ as follows
 \bea \label{EVEQST5} 
 S_{d + 2 d} \Lb Y\Rb \,\,&=&\, e^{- \,\Y}\sum^\infty_{j=0} \frac{e^{\gamma \,n\,j}\,\Y^j}{j!}  \frac{e^{ \gamma (n+1)}\,-\,1}{e^\gamma\,-\,1}\Lb -\,\gamma\,N_j\Rb^n n!\,\\
 &=&\,\,\,\frac{1}{\gamma}e^{- \,\Y}\sum^\infty_{j=0} \frac{\Y^j}{j!}\Bigg\{e^\gamma
\frac{1}{\gamma\,e^\gamma N_j} \exp\Lb\frac{1}{\gamma\,e^\gamma N_j}\Rb \,\Gamma\Lb 0,\frac{1}{\gamma e^\gamma N_j}\Rb\,-\, \frac{1}{\gamma\, N_j} \exp\Lb\frac{1}{\gamma\, N_j}\Rb \,\Gamma\Lb 0,\frac{1}{\gamma  N_j}\Rb \Bigg\}\nn\\
\xrightarrow{\Y \gg 1} & & e^{ - \tilde{\Delta}_1\,Y}
\nn
\eea
where $\Y \equiv \frac{\Delta}{\gamma} \,Y$.

Therefore, one can see that in the UTM \eq{EVEQST4} does not work. However, it turns out that we can get a simple formula for the Borel image of the scattering amplitude.  Indeed, as we have seen in \eq{SA304} , for $S\Lb Y \Rb$ we can write:
\beq \label{EVEQST6}
S_{d + d}\Lb \Y\Rb\,\,=\,\,
\,\,e^{- \,\Y}\sum^\infty_{j=0} \frac{\Y^j}{j!}\intl^\infty_0 d \tau e^{-\tau}
b^{(0)}_{d+d} \Lb \tau, N_j\Rb
~~~~~\mbox{with}~~~~~N_j\,\,=\,\,\gamma\,e^{\gamma\,j}
\eeq   

In the region of small $\gamma$ one can see from \eq{EVEQST3} that
\beq \label{EVEQST6}
S_{d+2d}\Lb \Y\Rb\,\,=\,\,
\,\,e^{- \,\Y}\sum^\infty_{j=0} \frac{\Y^j}{j!}\intl^\infty_0 d \tau e^{-\tau}
b^{(0)}_{d+2d} \Lb \tau, N_j\Rb~~~\mbox{with} ~~b^{(0)}_{d+2d} \Lb \tau, N_j\Rb\,\,=\,\,\, \frac{ d \Lb \tau N_j b^{(0)}_{d+d} \Lb \tau, N_j\Rb\Rb}{ d \Lb\tau N_j\Rb}\eeq

Using $ b^{(0)}_{d+d} \Lb \tau, N_j\Rb\,\,= 1/\Lb 1 \,\,+\,\, \tau\,N_J\Rb$ 
one can see that \eq{EVEQST1} for $ b^{(0)}_{d+d} \Lb \tau, N_j\Rb$ takes a form:
\beq \label{EVEQST7}
b^{(0)}_{d+d} \Lb \tau, N_{j+1}\Rb\,\,-\,\,b^{(0)}_{d+d} \Lb \tau, N_j\Rb = 
\Lb b^{(0)}_{d+d} \Lb \tau, N_j\Rb\Rb^2\,\,-\,\,b^{(0)}_{d+d} \Lb \tau, N_j\Rb
\eeq
Therefore, for the Borel images equation has the same form as \eq{EVEQST1} if we replace $\frac{d S}{d Y} $ by $b^{(0)}_{d+d} \Lb \tau, N_{j+1}\Rb\,\,-\,\,b^{(0)}_{d+d} \Lb \tau, N_j\Rb$ and $N_0 $ by $ 1 - b^{(0)}_{d+d}$.


       \subsubsection{ Single diffraction}


In Ref.\cite{KOLE}   it is shown that the equation for the cross section of the diffraction dissociation  for the BFKL cascade can be written in the form:

\beq \label{EVEQDD1}
\frac{d}{d \,\Delta\, Y} S^D_{d+d} \Lb Y \Rb \,\,=\,\, \Lb S^D_{d+d} \Lb Y \Rb \Rb^2\,\,-\,\, S^D_{d+d} \Lb Y \Rb ~~~\mbox{where}  ~~S^D \,=\,1- 2N_0 + \sigma_{sd}
\eeq

In the UTM it takes the following form:
\beq \label{EVEQDD2}
\frac{d}{d \,\Delta\, Y} S^D_{d+d} \Lb Y \Rb \,\,=\,\, \Lb S^D_{d+2d} \Lb Y \Rb \Rb\,\,-\,\, S^D_{d+d} \Lb Y \Rb ~~~\mbox{where}  ~~S^D \,=\,1- 2N_0 + \sigma_{sd}
\eeq
with the same reasoning as in \eq{EVEQST2}. For Borel images it translates to the equation:
\beq \label{EVEQDD2}
b^{(D)}_{d+d} \Lb \tau, N_{j+1}\Rb\,\,-\,\,b^{(D)}_{d+d} \Lb \tau, N_j\Rb = 
 b^{(D)}_{d+2d} \Lb \tau, N_j\Rb\,\,-\,\,b^{(D)}_{d+d} \Lb \tau, N_j\Rb= 
\Lb b^{(D)}_{d+d} \Lb \tau, N_j\Rb\Rb^2\,\,-\,\,b^{(D)}_{d+d} \Lb \tau, N_j\Rb\eeq
In the region of small $\gamma$ this equation can be rewritten  in the form of \eq{EVEQST2}, viz.:
\beq \label{EVEQDD3}
b^{(sd)}_{d+d} \Lb \tau, N_{j+1}\Rb\,\,-\,\,b^{(sd)}_{d+d} \Lb \tau, N_j\Rb\,\,=\,\,b^{(sd)}_{d+d} \Lb \tau, N_j\Rb\,+\,\Lb b^{(sd)}_{d+d} \Lb \tau, N_j\Rb\Rb^2\,-\,4\,b^{(0)}_{d+d}\Lb \tau, N_j\Rb\,
 b^{(sd)}_{d+d} \Lb \tau, N_j\Rb\,+\,\,2\,\Lb b^{(sd)}_{d+d} \Lb \tau, N_j\Rb\Rb^2
\eeq
  As have been discussed the AGK cutting rules give 
  \beq \label{EVEQDD4}
   b^{(sd)}_{d+d} \Lb \tau, N_j\Rb\,\,=\,\,\frac{ 2\,\Lb \tau\, \gamma\,  N_j\Rb^2}{\Lb 1\,\,+\,\,\tau\, \gamma\,N_j\Rb\,\Lb 1\,\,+\,\,2\,\tau\, \gamma\,N_j\Rb }
   \eeq  
   We obtain    \eq{EVEQDD3}    taking into account that
     \bea \label{EVEQDD5}   
     b^{(sd)}_{d+2d} \Lb \tau, N_j\Rb\,\, &=&\,\,\frac{d}{ d ( \tau \, \gamma\,N_j)} \Bigg(( \tau \, \gamma\,N_j) \underbrace{\Lb \frac{ 2\,\Lb \tau\,\gamma\,   N_j\Rb^2}{\Lb 1\,\,+\,\,\tau\, \gamma\,N_j\Rb\,\Lb 1\,\,+\,\,2\,\tau\, \gamma\,N_j\Rb } \Rb }_{    b^{(sd)}_{d+d} \Lb \tau, N_j\Rb}\Bigg)\,\,\nn\\
     &=&\,\,2   
 \,b^{(sd)}_{d+d} \Lb \tau, N_j\Rb \,-\,2\Lb  b^{(0)}_{d+d} \Lb \tau, N_j\Rb\Rb^2 + \Lb \frac{2 \tau \,N_j}{1\,+\,2\,\tau\, \gamma\,N_j}\Rb^2
 \eea   

     \begin{boldmath}
       \subsubsection{ $\sigma_1$}
  \end{boldmath}
  
 The equation for $\sigma_1$ takes the following form in the UTM:
\beq \label{EVEQS11}
\frac{d}{d \,\Delta\, Y} \sigma^{d+d}_1 \Lb Y \Rb \,\,=\,\, \sigma^{d+2d}_1 \Lb Y \Rb \,\,-\,\, \sigma^{d+d}_1 \Lb Y \Rb \eeq
This equation can be rewritten for the Borel images as follows:
\beq \label{EVEQS12}
b^{(1)}_{d+d} \Lb \tau, N_{j+1}\Rb\,\,-\,\,b^{(1)}_{d+d} \Lb \tau, N_j\Rb\,\,=\,\,b^{(1)}_{d+d} \Lb \tau, N_j\Rb\,+\,2\, b^{(1)}_{d+d} \Lb \tau, N_j\Rb\,b^{(sd)}_{d+d} \Lb \tau, N_j\Rb\,-\,4\,b^{(0)}_{d+d}\Lb \tau, N_j\Rb\,
 b^{(1)}_{d+d} \Lb \tau, N_j\Rb\eeq

Using the AGK prediction for $\sigma_1$
\beq \label{EVEQS13}
   b^{(1)}_{d+d} \Lb \tau, N_j\Rb\,\,=\,\,\frac{ 2\, \tau\, \gamma\,  N_j}{\Lb 1\,\,+\,\,2\,\tau\, \gamma\,N_j\Rb^2 }
   \eeq  
    we obtain
    \beq \label{EVEQS14}   
     b^{(1)}_{d+2d} \Lb \tau, N_j\Rb\,\, =\,\,\frac{d}{ d ( \tau \, \gamma\,N_j)} \Bigg(( \tau \, \gamma\,N_j) \underbrace{\Lb \frac{ 2\,\Lb \tau\,\gamma\,  N_j\Rb^2}{\Lb 1\,\,+\,\,2\,\tau\, \gamma\,N_j\Rb^2 } \Rb }_{    b^{(1)}_{d+d} \Lb \tau, N_j\Rb}\Bigg)\,
     =\,\,2   
 \,b^{(1)}_{d+d} \Lb \tau, N_j\Rb \,-\, \frac{4 \Lb \tau \,\gamma\,N_j\Rb^2}{\Lb1\,+\,2\,\tau\, \gamma\,N_j\Rb^3}
 \eeq     
   Plugging \eq{EVEQS14}, \eq{EVEQS13}  and \eq{EVEQDD4}  into \eq{EVEQS12} we see that this equation is correct.

   ~
   
   ~
   

     \begin{boldmath}
       \subsubsection{ $\sigma_n$}
  \end{boldmath}
  
 The equation for $\sigma_n$ has the same general form as
  \eq{EVEQS11}, viz.:

  \beq \label{EVEQSN1}
\frac{d}{d \,\Delta\, Y} \sigma^{d+d}_n \Lb Y \Rb \,\,=\,\, \sigma^{d+2d}_n \Lb Y \Rb \,\,-\,\, \sigma^{d+d}_n \Lb Y \Rb \eeq 
 
 Using that from AGK cutting rules the Borel image of $\sigma^{d+d}_n$ ( see \eq{XSK2})  is equal to
 \beq \label{EVEQSN2} 
b^{(n)}_{d+d}\Lb \tau, N_j\Rb\,\,=\,\, \frac{\Lb2\,\tau\,\gamma\,N_j\Rb^n}{\Lb1 \,\,+\,\,2\,\tau\gamma\,N_j\Rb^{n+1}} 
 \eeq
  and the Borel image of $ \sigma^{d+2d}_n$  can be written as follows
  \beq \label{EVEQSN3} 
 b^{(n)}_{d + 2d}\Lb \tau, N_j\Rb \,\,=\,\,\frac{d}{d\,N_j } \Lb N_j \,b^{(n)}_{d+d}\Lb \tau, N_j\Rb\Rb
 \eeq  
 we obtain the following equation for $b^{(n)}_{d+d}\Lb \tau, N_j\Rb $
 \bea \label{EVEQSN4}
&&b^{(n)}_{d+d} \Lb \tau, N_{j+1}\Rb\,\,-\,\,b^{(n)}_{d+d} \Lb \tau, N_j\Rb\,\,=\\&&\,\,b^{(n)}_{d+d} \Lb \tau, N_j\Rb\,+\,2\, b^{(n)}_{d+d} \Lb \tau, N_j\Rb\,b^{(sd)}_{d+d} \Lb \tau, N_j\Rb\,+\,\sum^{n-1}_{k=1}b^{(n-k)}_{d+d} \Lb \tau, N_j\Rb
b^{(k)}_{d+d} \Lb \tau, N_j\Rb-\,4\,b^{(0)}_{d+d}\Lb \tau, N_j\Rb\,
 b^{(n)}_{d+d} \Lb \tau, N_j\Rb\nn\eea
 
 In conclusion  we see that the general equation in the UTM cascade has the form of \eq{EVEQSN1} with $\sigma^{d + d}_0 \equiv\,\sigma_{sd}$. These equation are equivalent to the AGK cutting rules and have natural form with clear physics meaning. In the case of the BFKL cascade these equation can be reduced to \eq{EVEQ1} - \eq{EVEQ5} that are shown in \fig{eveq}.   In the region of small $\gamma$ which models QCD approach, we can write the similar equations but  for the Borel images of corresponding cross sections: see \eq{EVEQST7}, \eq{EVEQDD3}, \eq{EVEQS12} and \eq{EVEQSN4} .
 These equation has the same form as  \eq{EVEQ1} - \eq{EVEQ5} and can be written from them by replacing $\frac{d  }{d \Delta \,Y}\sigma_n\,\to\,b^{(n)}_{d + d}\Lb \tau,N_{j+1} \Rb\,-\,\,b^{(n)}_{d + d}\Lb \tau,N_{j} \Rb $ and $\sigma_n \,\to\, b^{(n)}_{d + d}\Lb \tau,N_{j} \Rb$.

 All equations for the BFKL cascade can be derived from the equation
\beq  \label{EVEQSN5} 
 \frac{d}{d \,\Delta\, Y}\underbrace{ S_{in} \Lb Y \Rb}_{select\,fixed \,multiplicity \,terms}
  \,\,=\,\,\underbrace{  \Lb S_{in} \Lb Y \Rb \Rb^2\,\,-\,\, S_{in} \Lb Y \Rb}_{ select\,fixed \,multiplicity \,terms}~~~\mbox{where}  ~~S_{in} \,=\,1- 2\,N_0 + \sigma_{sd}\,+\sum_{n=1} \sigma_{n}
  \eeq
 while for the UTM cascade \eq{EVEQSN5}  takes the form: 
  \beq  \label{EVEQSN6} 
\underbrace{ b_{in} \Lb \tau, N_{j+1}\Rb -  b_{in} \Lb \tau, N_{j}\Rb }_{select\,fixed \,multiplicity \,terms}
  \,\,=\,\,\underbrace{  \Lb  b_{in} \Lb \tau, N_{j}\Rb \Rb^2\,\,-\,\, b_{in} \Lb \tau, N_{j}\Rb}_{ select\,fixed \,multiplicity \,terms}~~~\mbox{where}  ~~b_{in} \,=\,1- 2\,b^{(0)}  + b^{(sd)}\,+\sum_{n=1} b^{(n)}
  \eeq  
  
 ~
 
 ~
 
 ~

    
    \section{Entropy of produced particles}   
    \begin{boldmath}
  \subsection{ Continuous approximation}
  \end{boldmath}  
  
 
 Here we will focus on the entropy of produced particles in the regime of  large rapidities  but $\gamma\,N(Y)\gg 1$ in the continuous approximation
 from \eq{AGKN1}, 
 \beq \label{ETM1} 
\displaystyle{ \sigma_n \,\,=\,\, \frac{1}{\gamma\,N} \,K_0\Bigg( 2 \sqrt{ \frac{n+1}{2\gamma\,N}}\Bigg)} 
 \eeq
 which is the probability to find $n$ particles in the final state. 
 In fact  $\sigma_n/\sigma_{in}$,  should be convoluted with  
 ${\cal P}^{\mbox{\tiny Pom}}_n$  (see \eq{AGKN2}), but at high energies, the latter can be approximated by a $\delta$-function
which  plays no role in computation of the entropy. 

\begin{figure}[ht]
    \centering
  \leavevmode
  \includegraphics[width=8cm]{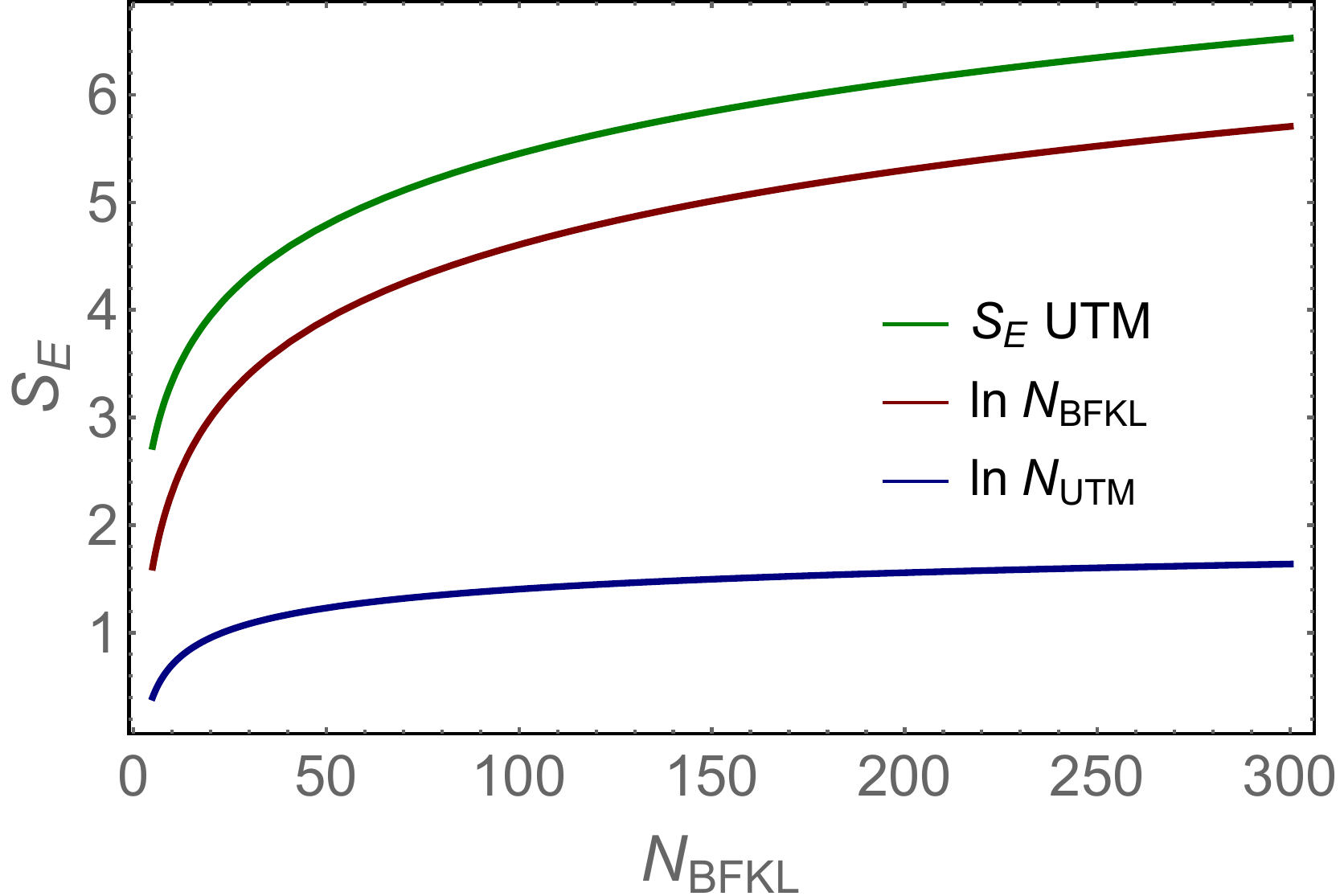}  
     \caption{Entropy comparison.  $S_E$ computed from the exact multiplicity distribution (\ref{ETM4}) in UTM  versus $\ln [N_{UTM/BFKL}]$ .    
     $S^{BFKL}_E = \ln N_{BFKL}$, where $N_{BFKL}$ is the mean multiplicity of produced  dipoles in DIS .  $N_{UTM}$ is the mean multiplicity of dipoles    in the initial state.
     $\Delta=0.2$ and $\gamma = 0.025$. }
\label{se}
   \end{figure}

The von Neumann entropy of the produced particles is:
\beq \label{ETM3}
S_E\,\,=\,\,- \sum_n \ln[ \sigma_n/\sigma_{in}] \,\,\frac{\sigma_n}{\sigma_{in} }
\eeq
Substituting \eq{ETM1} into \eq{ETM3} and switching from summation over $n$ to integration over 
$\eta = 2 \sqrt{\frac{n+1}{2\,\gamma\,N}}$, we obtain
\beq \label{ETM4}
S_E\,\,=\,\,  \ln \Lb2 \gamma \,N\Rb\,\,-\,\,\int _0^\infty \eta\, d \eta \ln[ K_0\Lb \eta\Rb ]\,\, K_0\Lb \eta\Rb 
\eeq
 The integral over $\eta$ equals $-1.5$ which does not depend on $Y$.  Therefore, at large $Y$,
 \beq \label{ETM5}
S^{BFKL}_E\,\,=\,\,  \ln [ N_{BFKL}\Lb Y\Rb]\,\,+\,\,1.5  \,\,\xrightarrow{Y \gg 1}\,\,\Delta\,Y\,;
~~~~~~~~~~~N_{BFKL}=2\,\gamma\,N(Y)
\eeq
The result  for the entropy coincides with the result of Ref.\cite{KHLE} but  is quite different from \eq{I1}. 
It is important to mention that \eq{ETM5} is different from (see Fig. \ref{se}) 
\beq \label{ETM6}
S^{UTM}_E\,\,=\,\,  \ln N_{\mbox{\tiny UTM}}\,;~~~~~~~~~~~~N_{\mbox{\tiny UTM}}=\sum_n n\,P^{UTM}_n
\eeq

It should be emphasize that \eq{ETM4} has a general origin: it stems from the KNO scaling behaviour\cite{KNO}  of the multiplicity distribution\cite{utmm}. Indeed, if 
$\sigma_n/\sigma_{in} $ has KNO behavior, viz.:
\beq \label{ETM7}
\frac{\sigma_n}{\sigma_{in}} = \frac{1}{\bar n} \Psi\Lb \frac{n}{\bar n}\Rb
\eeq
where $ \bar n $ is a mean multiplicity, one can see that the entropy
\beq \label{ETM8}
S_E=\-\sum_n \ln \Lb \frac{1}{\bar n} \Psi\Lb \frac{n}{\bar n}\Rb\Rb \frac{1}{\bar n} \Psi\Lb \frac{n}{\bar n}\Rb=\ln \bar{n} + \sum \ln \Lb \Psi\Lb \frac{n}{\bar n}\Rb\Rb \frac{1}{\bar n} \Psi\Lb \frac{n}{\bar n}\Rb=\ln \bar{n} + \int d \zeta  \ln\Lb \Psi\Lb \zeta\Rb\Rb \Psi\Lb \zeta\Rb
\eeq
where $ \zeta = n/\bar{n}$. The integral over $\zeta$ leads to the contribution which does not depend on $Y$, while $\ln \bar{n} = \Delta Y$.

\begin{figure}[ht]
    \centering
  \leavevmode
  \begin{tabular}{c c }
  \includegraphics[width=8cm]{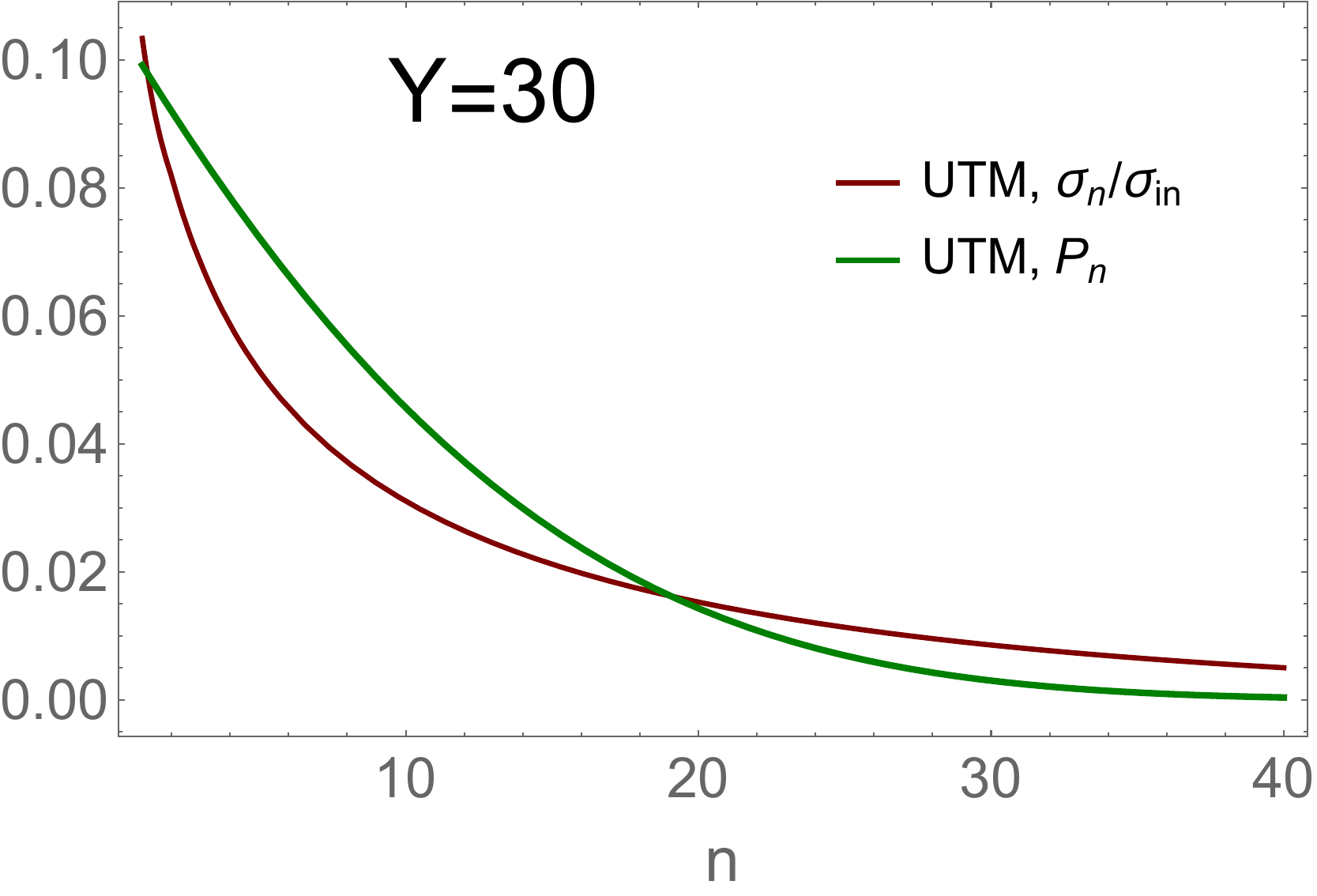}&  \includegraphics[width=8.3cm]{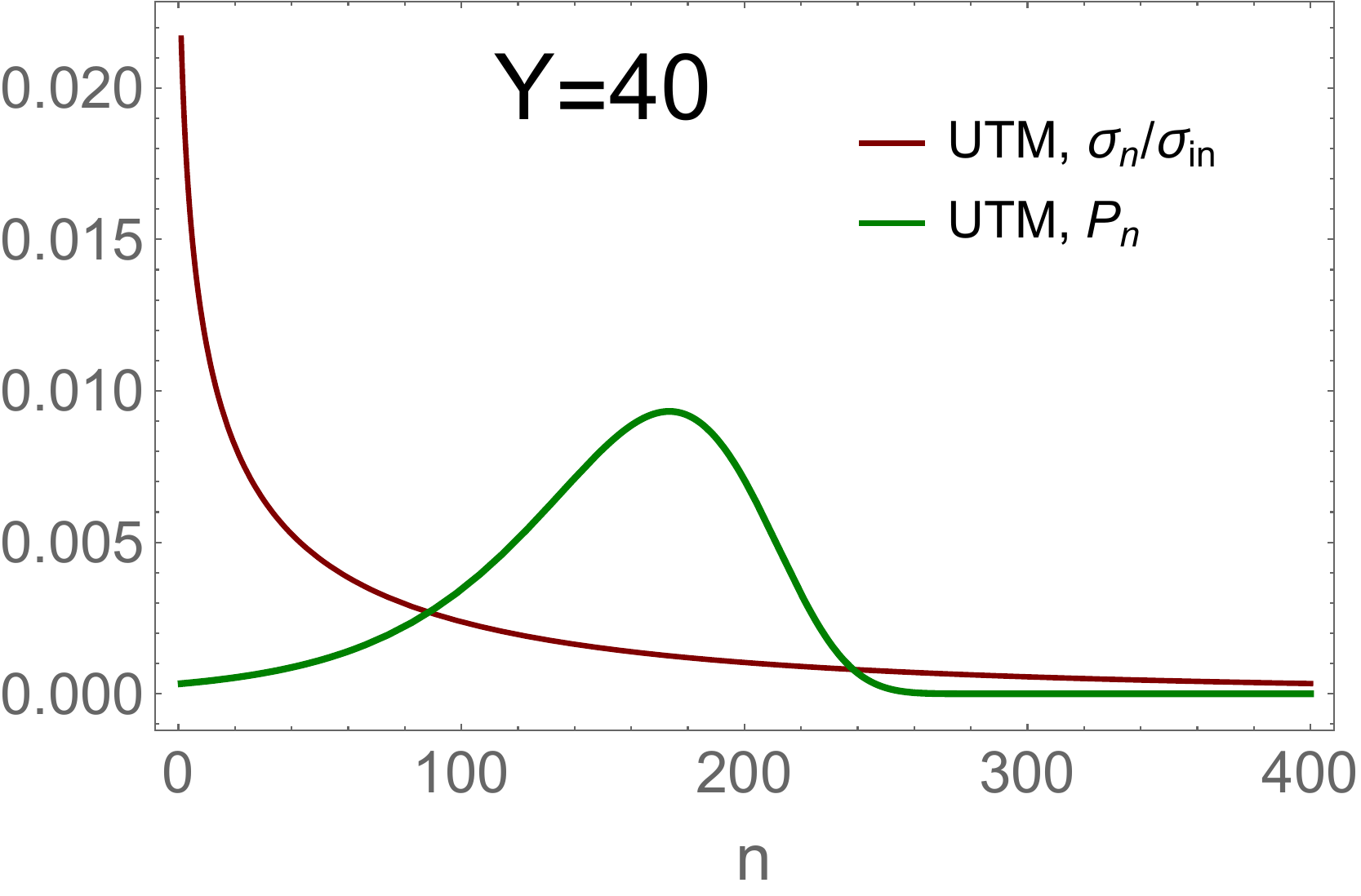}   \\
\end{tabular}
     \caption{${\sigma_n/\sigma_{in}} $  from    \eq{AGKN}  and $P_n=P_n^{UTM}$  for the parton cascade (see Ref.\cite{utmm})   versus $n$ at different $Y$. $\Delta=0.2$ and $\gamma = 0.025$. }
\label{utmcas}
   \end{figure}

What is striking that while the entropy (\ref{ETM3}) $\sim \ln N_{BFKL}$ , which is the result one would obtain for the 
BFKL  cascade in DIS, the underlying multiplicity distributions  for the particle production and for the parton cascade 
are quite different as illustrated in  \fig{utmcas}. Another explanation is that  despite the same entropy of the produced particles (see \eq{ETM8}) and the entropy of the parton in the wave functions \cite{utmm}
  the shape of distribution of the produced particles is quite different  from $P_n$.


\begin{figure}[ht]
    \centering
  \leavevmode
\includegraphics[width=10cm]{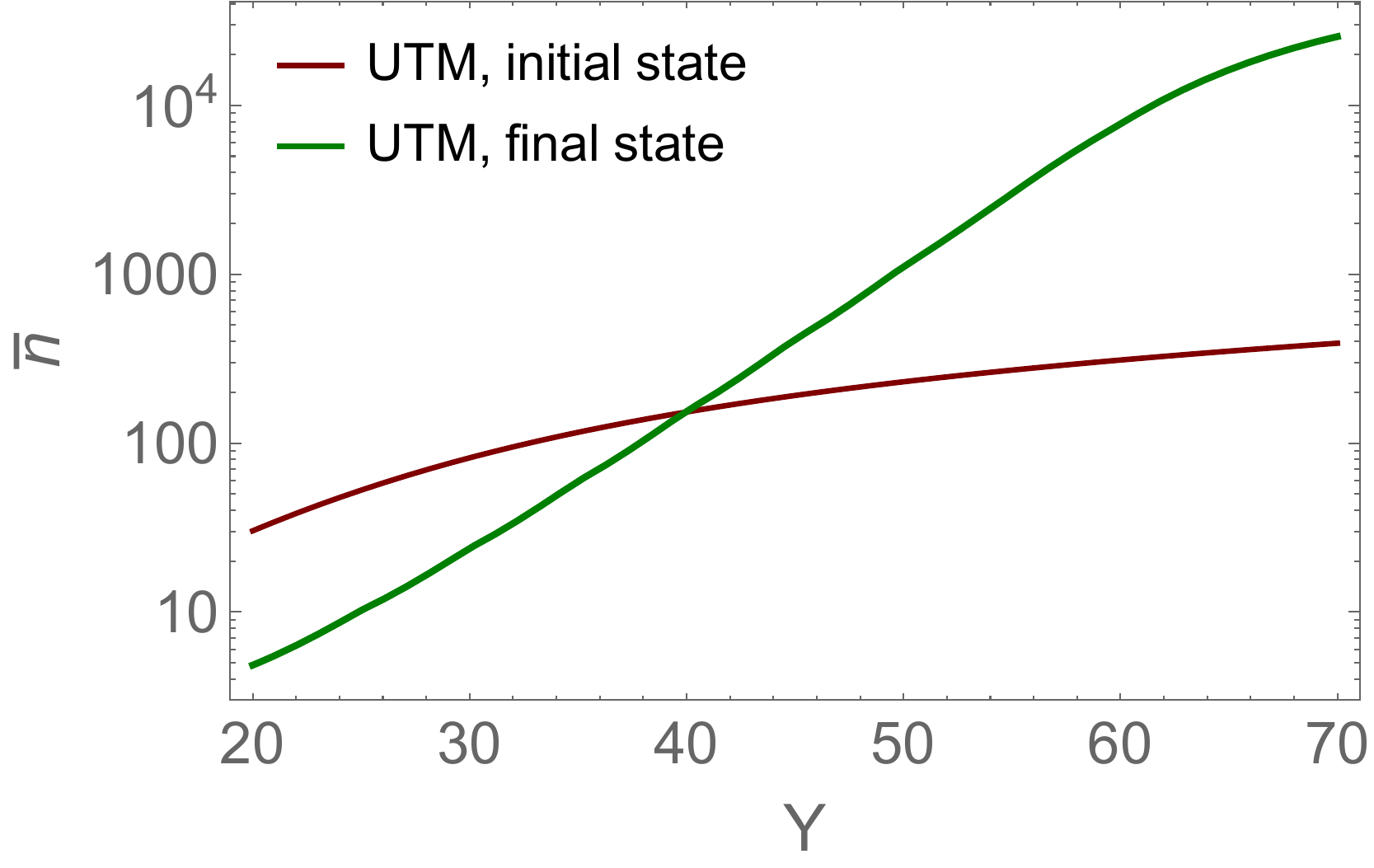}  \\
    \caption{The mean multiplicities in the UTM versus Y. $\Delta=0.2$ and $\gamma = 0.025$. }
\label{ifm}
   \end{figure}

In \fig{ifm} we plot the mean multiplicities for the initial and final states in the UTM. One can see that at large Y the much larger number of dipoles has been produced that it has the initial wave function.

  \begin{boldmath}
  \subsection{ Arbitrary $Y$}
  \end{boldmath}
 As has been discussed at large $Y$ we can use \eq{XSK4} for  $\sigma_n/\sigma_{in}$ . This equation leads to the expression for the entropy of the same form as \eq{ETM4} but the integral over $\eta$ has a more complex form being a function of $\Y$.
  It is equal to
  \bea \label{ETM9}
 \displaystyle{ I\Lb \Y\Rb= \frac{\int d \eta \Lb  2 \sqrt{\eta} + \Delta \gamma\,Y \eta\Rb\exp\Lb -  2 \sqrt{\eta} - \Delta \gamma\,Y \eta\Rb}{\int d \eta \exp\Lb -  2 \sqrt{\eta} - \Delta \gamma\,Y \eta\Rb} \,\,\rightarrow\,\,
 \left\{\begin{array}{l}\,\,2\,\,\,\,\,\,\,\,\,\,\mbox{for}\,\,\,\Delta \gamma Y \,\ll\,1;\\ \\
\,\,\,\,1\,\,\,\,\,\mbox{for}\,\,\,\Delta \gamma Y \,\gg\,1;\\  \end{array}
\right.}
 \eea
 
 Therefore, we can conclude that at large Y the entropy of produced dipoles is given by
 \eq{ETM5}: $ S_E \,\,=\,\,\Delta\,Y$. It should be noted that the same entropy was computed in the zero dimension model at $\tau-0$ in \fig{pc} \cite{KHLE}. Hence, in spite of the mess of interactions it turns out  that the entropy of the parton cascade  is not changed during its propagation  from $\tau=0$ to $\tau=\infty$.

 ~

     \begin{boldmath}
     \section{Conclusions }
      \end{boldmath}   
        
      
      In this paper we found the multiplicity distribution of the produced dipoles in the final state  for dipole-dipole scattering. This distribution show the great difference 
       from the distributions of the parton in the wave function of the projectile. However, in spite  of this difference the entropy of the produced dipoles turns out to be the same as the entropy of the dipoles in the wave function.  This fact is not surprising since we have discussed in the paper that in the parton approach only dipoles in the hadron wave function can be produced at $t = +\infty$ and measured by the detectors.
We also confirm the result of Ref.\cite{KHLE} that this entropy is equal to $S_E = \ln\Lb xG(x)\Rb$, where we denote by $xG$ the mean multiplicity of the dipoles in  the deep inelastic scattering. This result  contradicts    the CGC-Black Hole correspondence , that has been suggested in Ref.\cite{DVVE}.  

Our results are based on AGK cutting rules but we derived the evolution equations for the partial cross sections $\sigma_n$ which have the clear physics and reproduce the same results as the AGK cutting rules.

       It should be stressed that we are able to consider in the zero dimension models not only DIS but the hadron-hadron scattering.  The above result can be reformulated as the statement that the average multiplicity of produced dipoles coincides with mean multiplicity of the DIS for hadron-hadron collisions.

      Second,  we demonstrated a different mechanism for high energy multiparticle generation process ,which naturally arises in  the zero dimension models   from summing large Pomeron loops, and  in which we do not need to produce a classical state with maximal entropy to satisfy $t$ and $s$ channel unitarity.   We believe that this mechanism will work in QCD. 
 
     ~
     
     ~

       {\bf Acknowledgements} 
     
   We thank our colleagues at Tel Aviv university for
 discussions. Our  special thanks goes to Alex Kovner and Michael Lublinsky for  the encouraging  skepticism  about  AGK cutting rules which results in the appearance of section IV in the paper.
This research  was funded by Binational Science Foundation grant 2021789.

 ~
 
 ~
  
 \appendix
\section{Distribution of the produced gluons in the BFKL Pomeron}

 Here we wish to stress that this follows from the general structure of the BFKL Pomeron in QCD\cite{BFKL}.   Indeed, the BFKL equation has the general form in momentum representation:
  \beq \label{FS1}
  \frac{\partial\,\phi_n\Lb Y,  k\Rb }{\partial\,Y}\,\,=\,\,\int d^2 k'\,K\Lb k,k'\Rb \,\phi_{n-1}\Lb Y,k'\Rb
  \eeq
  where $\phi_n$ is the cross section of $n$ produced gluons and $K(k,k')$ is the kernel (see Ref.\cite{BFKL} for details). The eigenfunction of this equation are $\Lb k^2\Rb^\gamma$  and the general solution is the sum over these eigenfunctions
  \beq \label{FS2}
  \phi_n\Lb Y, k\Rb\,\,=\,\,\sum_\gamma C\Lb Y, n\Rb\,\Lb k^2\Rb^\gamma  
   \eeq
   
   with the following equation for coeifficients $C\Lb Y, n\Rb$:
    \beq \label{FS3}
 \frac{\partial C\Lb Y, n\Rb}{\partial\,Y} \,\,=\,\,\chi\Lb \gamma\Rb C\Lb Y, n-1\Rb 
   \eeq   
   where $\chi$ is the image of the BFKL kernel. One can check that the solution to \eq{FS3} takes the form:
     \beq \label{FS4}
C\Lb Y, n\Rb\,\,=\,\,\frac{\Lb \chi(\gamma)\,Y\Rb^n}{n!}\eeq    
  leading to
  \beq \label{FS5}
  \phi_n\Lb Y, k\Rb\,\,=\,\,\int \frac{ d \gamma}{2 \pi i} \frac{\Lb \chi(\gamma)\,Y\Rb^n}{n!}\Lb k^2\Rb^\gamma  
   \eeq   
   The cross section is equal to
     \beq \label{FS6}
 \sigma_{tot}\,=\,\sum_{n=0}^\infty \phi_n\Lb Y, k\Rb\,\,=\,\,\int \frac{ d \gamma}{2 \pi i}e^{\chi\Lb \gamma\Rb \,Y} \Lb k^2\Rb^\gamma  
   \eeq     
   while the probability to find $n$ gluons in the final state is equal to
    \beq \label{FS7}
P_n\,=\, \frac{\phi_n\Lb Y, k\Rb}{\sigma_{tot}} \,\,=\,\,\frac{\int \frac{ d \gamma}{2 \pi i} \frac{\Lb \chi(\gamma)\,Y\Rb^n}{n!}\Lb k^2\Rb^\gamma}{  \int \frac{ d \gamma}{2 \pi i}e^{\chi\Lb \gamma\Rb \,Y} \Lb k^2\Rb^\gamma } 
   \eeq    
   At high energies the main contribution in \eq{FS7} can be estimated using the method of steepest descent with $\gamma_{SP} = \h$. Bearing this in mind we see that \eq{FS7} give the Poisson distribution.
   
\section{The Pomeron calculus of the UTM}


In this appendix we wish to review what we know about the Pomeron calculus in zero dimension models and, particularly  in the UTM. We have to admit that the majority of features of  this calculus have been known for long time (see Refs.\cite{ACJ,AAJ,JEN,ABMC,CLR,CIAF,MUSA,nestor,RS,SHXI,KOLEV,BIT,LEPRI,utm,utmm,utmp})  but it turns out that a substantial part of these properties was preserved in my memory and I am sorry for not finding a proper references.

 In our paper\cite{utm} we showed that the UTM model can be viewed as the Pomeron calculus with the following Hamiltonian:

 \beq \label{TAK4}
\mathcal{H}_{\rm UTM}\,\,=\,\,-\bar{\pom} \,\pom
 \eeq
 where $\pom$ and $\bar{\pom}$ are Pomeron operators, generates the rich Pomeron calculus. In \eq{TAK4} we introduce $\Y = \,\frac{\Delta}{\gamma} Y$. In the notations of this paper the Pomeron Hamiltonian  is shown in \eq{ZEQG} and it has a form:
  \beq \label{TAK4}
\mathcal{H}_{\rm UTM}\,\,=\,\, 
- t\,\Lb1-e^{\gamma (1 - t)\frac{\partial}{\partial t}}\Rb\,\,=\,\,
 \underbrace{ \gamma t \frac{\partial}{\partial t}}_{ \mathcal{H}_0} +  
 \underbrace{ t\,\Lb e^{\gamma (1 - t)\frac{\partial}{\partial t}}\,\,-\,\,1\,\,-\,\,\gamma\frac{\partial}{\partial\,t}\Rb}_{ \mathcal{H}_I} 
 \eeq 
 where $u = 1-t$ (see Ref.\cite{LELU} for details). The choice of $\mathcal{H}_0$ is arbitrary and we did it based on the interpretation of the zero dimension models as the QCD with fixed dipole size. We have discussed this interpretation in the introduction (see \eq{I1}).
This calculus have all vertices for $n \pom \to m \pom$ ($V^n_m$)  interactions, in spite the evolution equation for $P_n$ has only two vertices due to decay of one dipole to two (see \eq{MEQ}).
It also gives the first  and the only example of solvable simple Pomeron calculus with the infinite number of vertices. Expanding $\mathcal{H}_I$ with respect to small $\gamma$ one can see that in the order of $\gamma$ we have the triple Pomeron vertex $ \pom \to 2 \pom$ ($V^1_2$) which is equal to $ - \gamma$. In the order of $\gamma^2$ $\mathcal{H}_I$ generates
the correction to $V^1_1 = - \h \gamma^2 (\pom \to \pom)$ and vertices
s $2 \pom \to \pom, \, 2 \pom\to 2 \pom$ and $  2 \pom\to 3\pom $. They are equal to
 $V^2_1 =\h \gamma^2$ , $ V^2_2 = -\gamma^2$ and $V^2_3 = \h \gamma^2$ (see \fig{diapr}). 

\subsection{The first Pomeron diagrams}

\begin{figure}[ht]
    \centering
  \leavevmode
      \includegraphics[width=16cm]{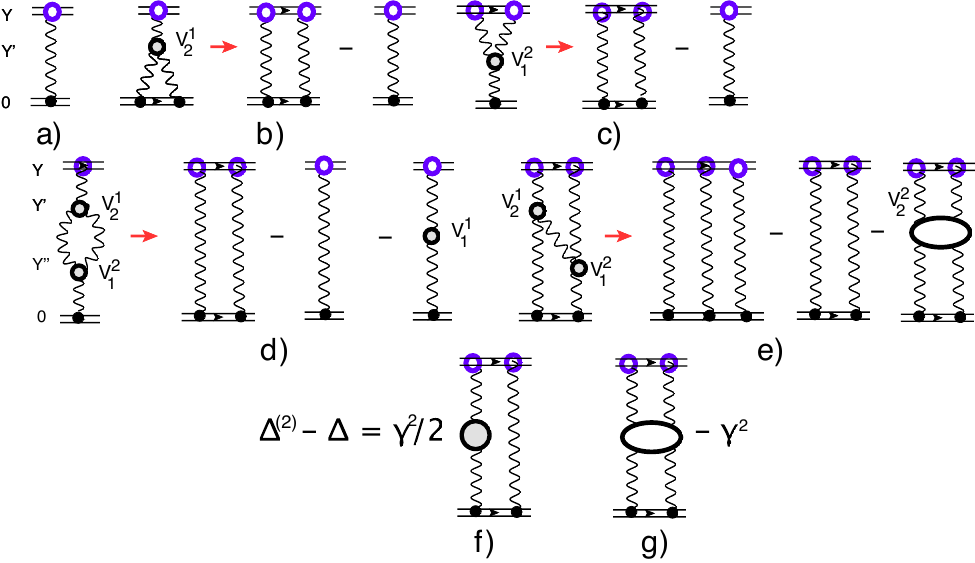}  
      \caption{ The first Pomeron diagrams for $Z\Lb u, Y\Rb$ .  \fig{diapr}-a: the exchange of the Pomeron. \fig{diapr}-b and \fig{diapr}-c: the   first semi-enhanced diagrams.  \fig{diapr}-d:  the diagrams for the  Pomeron Green's function.\fig{diapr}-e the first diagrams for the two Pomeron Green's function. The black circle is the vertex of interaction of the Pomeron with the target, which is equal to $\gamma\,\frac{d}{d\,t} $. The  blue circle is the vertex of interaction with the projectile dipole: -$t$.  $V^m_n$ are vertices of the transitions $m \pom \to n\pom$. $V^1_2 = \gamma t^2 \frac{d}{d t} $, $V^2_1 = \h \gamma^2 t \frac{d^2}{d t^2}$,$V^1_1= - \h \gamma^2 t \frac{d}{d t}$, and 
 $V^2_2 $ is vertex for $2\pom \to 2 \pom$ interaction, which is equal to $ \h \gamma^2 t^2 \frac{d^2}{d t^2}$.}
\label{diapr}
   \end{figure}

The main features of the Pomeron calculus in the UTM can be seen just from the first diagrams (see \fig{diapr}). We will calculate them in $\omega$ representation, which is defined as
\beq \label{PC1}
Z\Lb u=1-t \,Y\Rb\,\,=\,\,\int\limits^{\epsilon + i \infty}_{\epsilon - i \infty} \frac{ d \,\omega}{2\,\pi\,i} \,e^{ \omega\,\Y}\,z\Lb t,  \omega\Rb\eeq
 
 \eq{ZEQG} in this representation has the following form:
 \beq \label{PC2} 
 \mathcal{H}_{\rm UTM} \,z\Lb t , \omega\Rb =  \Lb \mathcal{H}_0\,\,+\,\,\mathcal{H}_I \Rb\,z\Lb t,\,  \omega\Rb  \,\,=\,\,\omega\,z\Lb t,\,  \omega\Rb 
 \eeq
  The first diagram of \fig{diapr}-a is the exchange of the 'bare' Pomeron, which gives the contribution
   \beq \label{PC3} 
   z\Lb t =\gamma,\,  \omega; \fig{diapr}-a\Rb  \,\,=\,\,\frac{\gamma}{\omega - \Delta}~~~~\mbox{with}~~~ \Delta = \gamma 
   \eeq
   where $\Delta$ is the intercept of the 'bare' Pomeron.
   
  The contribution of the diagram of \fig{diapr}-b takes the form 
   \beq \label{PC4} 
   z\Lb t =\gamma,\,  \omega; \fig{diapr}-b\Rb  \,\,=\,\,-\frac{ V^1_2\,\gamma^2}{\Lb\omega - \Delta\Rb \Lb\omega - 2 \Delta\Rb}\, =- \frac{ V^1_2\,\gamma^2}{\Delta}\Bigg( \underbrace{ \frac{1}{\omega - 2 \Delta}}_{2 \pom} \,-\, \underbrace{ \frac{1}{\omega -  \Delta}}_{ \pom}\Bigg) =
   - \frac{\gamma^2}{\omega - 2 \Delta}\,+\, \frac{\gamma^2}{\omega - \Delta}  
 \eeq 
 
 From \eq{PC4} one can see that the semi-enhanced diagram of \fig{diapr}-b
 give the new  contribution, which is the exchange of two 'bare' Pomerons and the corrections to the vertex of the 'bare' Pomeron interaction with the target (see \fig{diapr}-a), which becomes $\gamma + \gamma^2$.
 
 The diagram of \fig{diapr}- gives the same contribution.
 Indeed,
  \beq \label{PC5} 
   z\Lb t =\gamma,\,  \omega; \fig{diapr}-c\Rb  \,\,=\,\,-\frac{ 2\,V^2_1\,\gamma}{\Lb\omega - \Delta\Rb \Lb\omega - 2 \Delta\Rb}\, =- \frac{ 2\,V^2_1\,\gamma}{\Delta}\Bigg( \underbrace{ \frac{1}{\omega - 2 \Delta}}_{2 \pom} \,-\, \underbrace{ \frac{1}{\omega -  \Delta}}_{ \pom}\Bigg) =
   - \frac{\gamma^2}{\omega - 2 \Delta}\,+\, \frac{\gamma^2}{\omega - \Delta}  
 \eeq 
 
 For $V^2_1$ we use $\h  \gamma \frac{d^2}{d t^2}$ as has been discussed above. It is instructive to note that in the BFKL cascade we do not have the diagram of \fig{diapr}-c  and, therefore,  the contribution of the two 'bare' Pomeron exchange is in two times less than in the UTM.  This feature we can see directly from \eq{SA304}.
 
 Now we calculate the first enhanced diagram of \fig{diapr}-d, whose contribution is equal to
 
   \bea \label{PC6} 
   z\Lb t =\gamma,\,  \omega; \fig{diapr}-d\Rb &=&-\frac{V^1_2 2\,V^2_1\,\gamma}{\Lb\omega - \Delta\Rb \Lb\omega - 2 \Delta\Rb \Lb \omega - \Delta\Rb}  = -\frac{V^1_2 2\,V^2_1\,\gamma}{\Lb\omega - \Delta\Rb} \frac{1}{\Delta} \Bigg( \frac{1}{ \omega - 2 \Delta} - \frac{1}{ \omega - \Delta}\Bigg) \nn\\
   &=& -\frac{V^1_2 2\,V^2_1\,\gamma}{\Delta} \Bigg( \frac{1}{\Delta} \Lb \underbrace{\frac{1}{\omega - 2 \Delta} }_{2 \pom} - \underbrace{\frac{1}{\omega -  \Delta} }_{\pom}\Rb -\underbrace{\frac{1}{\Lb\omega -  \Delta\Rb^2} }_{\mbox{intercept}\,\, \pom}\Bigg)\nn\\
   &=&- \gamma^2 \Lb\frac{1}{\omega - 2 \Delta}  \,-\,\frac{1}{\omega - \Delta}
   \Rb\,+\, \fbox{$\gamma^2$} \gamma \frac{1}{\Lb\omega -  \Delta\Rb^2} \eea
 The four diagrams that we have  calculated exhaust the contribution of the order of $\gamma^2$ to the Green's function of the Pomeron. Collecting terms proportional to $1/\Lb \omega - \Delta\Rb$ we have the following contributions:
   \beq \label{PC7}  
 G_{\pom}\Lb \omega\Rb= \Lb \gamma + \frac{3}{2} \gamma^2\Rb\frac{1}{\omega - \Delta} \,\,- \,\,\gamma \frac{\h\gamma^2} {\Lb \omega - \Delta \Rb^2}\eeq
 Note that in the second term we took into account the contribution of $V^1_1$ in the order of $\gamma^2$. It is easy to see that \eq{PC7} reproduce \eq{SA304} if we expand both the Pomeron intercept and the impact factor to the order $\gamma^2$. 
 
 Now we are going to evaluate the intercept of the two Pomeron Green's function. First contribution come from the exchange of two non-interacting Pomerons with a new intercept  of the order of $\gamma^2$(see \fig{diapr}-f): $\Delta^{(2)} = 
\Delta \,+\,\h \gamma^2  =   \gamma \,+\,\h \gamma^2$, as we have found in \eq{PC7}. The second contribution stems from \fig{diapr}-e. Indeed, this diagram is equal to 
 \bea \label{PC8} 
   z\Lb t =\gamma,\,  \omega; \fig{diapr}-e\Rb &=&2\frac{V^1_2 2\,V^2_1\,\gamma^2}{\Lb\omega -  2 \Delta\Rb \Lb\omega - 3 \Delta\Rb \Lb \omega - 2 \Delta\Rb}  = 2\frac{V^1_2 2\,V^2_1\,\gamma^2}{\Lb\omega -  2 \Delta\Rb} \frac{1}{\Delta} \Bigg( \frac{1}{ \omega - 3 \Delta} - \frac{1}{ \omega -  2\Delta}\Bigg) \nn\\
   &=& 2\frac{V^1_2 2\,V^2_1\,\gamma^2}{\Delta} \Bigg( \frac{1}{\Delta} \Lb \underbrace{\frac{1}{\omega - 3 \Delta} }_{3 \pom} - \underbrace{\frac{1}{\omega - 2 \Delta} }_{2 \pom}\Rb -\underbrace{\frac{1}{\Lb\omega -  2 \Delta\Rb^2} }_{\mbox{intercept}\,\,2 \pom}\Bigg)\nn\\
   &=&\gamma^3 \Lb\frac{1}{\omega - 3 \Delta}  \,-\,\frac{1}{\omega - 2 \Delta}
   \Rb\,-\, \fbox{$2 \gamma^2$} \gamma^2 \frac{1}{\Lb\omega -  2 \Delta\Rb^2} \eea
 In addition to \fig{diapr}-e and \fig{diapr}-f we need to calculate the diagram of \fig{diapr}-g, in which we take into account the vertex  $V^2_2 = -\gamma^2$  for direct $ 2 \pom \to 2 \pom$ transition. Sum of all these contributions gives the intercept of $G_{2 \pom}$ $\Delta_2^{(2)} = 2 \gamma + 2 \gamma^2$ in accord with \eq{SA304}. 
 
 ~

\subsection{Summing enhanced diagrams}

\begin{figure}[ht]
    \centering
  \leavevmode
      \includegraphics[width=16cm]{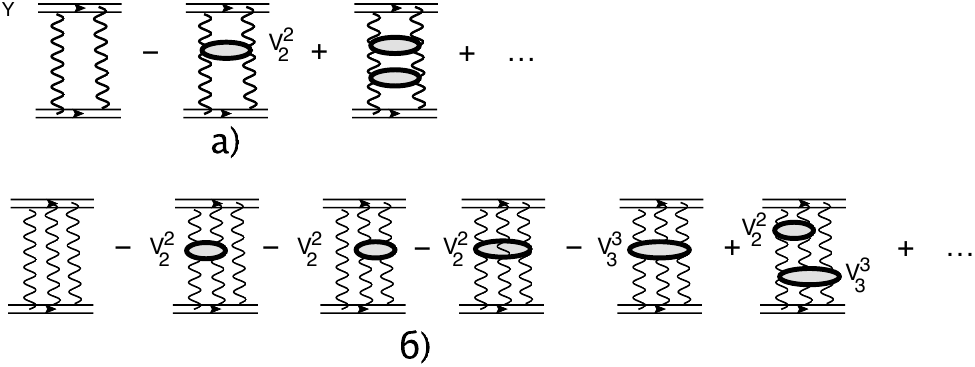}  
      \caption{ The typical Pomeron diagrams in the UTM.   \fig{dia}-a: the diagrams for two Pomeron Green's function. \fig{dia}-b:  the diagrams for the three Pomeron Green's function.  $V^2_2 $ is vertex for $2\pom \to 2 \pom$ interaction, while $V^3_3 $  is vertex for $3\pom \to 3 \pom$ interaction .The vertices $V^n_n $ are equal to $\Lb e^\gamma - 1\Rb^n$.}
\label{dia}
   \end{figure}

In this section we demonstrate that the intercepts for the Pomeron Green's function from \eq{SA304}  can be derived for the vertices $V^n_n = \Lb e^\gamma\,-\,1\Rb^n$ (see \fig{dia}).  The simplest case is $G_{2 \pom}$ which is shown in \fig{dia}-a. The bold wavy lines denote the Pomeron with the intercept $\Delta_1 = e^\gamma - 1$. The sum of diagrams of \fig{dia}-a has the following form
\beq \label{PC9}
G_{2 \pom}\Lb \omega\Rb = - \frac{1}{\omega - 2 \Delta_1} - \frac{1}{\omega -  2\Delta_1} V^2_2 \frac{1}{\omega - 2 \Delta_1} - \frac{1}{\omega - 2 \Delta_1} V^2_2 \frac{1}{\omega - 2 \Delta_1} V^2_2 \frac{1}{\omega - 2\Delta_1} - \dots = -\frac{1}{\omega - 2\,\Delta_1 - V^2_2}
\eeq
%
Plugging $V^2_2 = \Lb e^\gamma\,-\,1\Rb^2$ in \eq{PC9} we obtain
\beq \label{PC10}
G_{2 \pom}\Lb \omega\Rb =- \frac{1}{\omega - 2\Lb e^\gamma \,-\,1\Rb -\Lb e^\gamma \,-\,1\Rb^2} = - \frac{1}{\omega - \Delta_2} = -\frac{1}{\omega - \Lb e^{2 \gamma}  \,-\,1\Rb}
\eeq
Hence \eq{PC10} reproduces the exchange of two Pomerons in \eq{SA304}.

For $G_{3 \pom} \Lb \omega\Rb$ of \fig{dia}-b we have
\bea \label{PC11}
G_{3 \pom}\Lb \omega\Rb &=&  
\frac{1}{\omega - 3 \Delta_1} +\frac{1}{\omega -  3\Delta_1}\Lb 3\, V^2_2  + V^3_3\Rb\frac{1}{\omega - 3 \Delta_1} \\
&+& \frac{1}{\omega - 3 \Delta_1}\Lb 3\, V^2_2  + V^3_3\Rb\frac{1}{\omega - 3 \Delta_1}  \Lb 3\, V^2_2  + V^3_3\Rb \frac{1}{\omega - 3\Delta_1} +\dots = -\frac{1}{\omega - 3\,\Delta_1 - 3  V^2_2  - V^3_3}\nn\eea
Taking into account that $V^2_2= \Lb e^\gamma - 1\Rb^2$ and $V^3_3=\Lb e^\gamma - 1\Rb^3$ we have
\beq \label{PC12}
G_{ \pom}\Lb \omega\Rb =- \frac{1}{\omega - 3\Lb e^\gamma \,-\,1\Rb - 3\,\Lb e^\gamma \,-\,1\Rb^2  - \Lb e^\gamma \,-\,1\Rb^3} = - \frac{1}{\omega - \Delta_2} = -\frac{1}{\omega - \Lb e^{3 \gamma}  \,-\,1\Rb}
\eeq
which reproduces the intercept of the third term in \eq{SA304}. It is easy to generalize the derivation to $G_{n \pom}\Lb \omega \Rb$.

 Therefore,  the scattering amplitude of \eq{SA1} has simple and apparent meaning: each term is the Green's function for the exchange of $n$-Pomerons which is given by the factor $\exp\Lb \Delta_n \,\Y\Rb$ (see \fig{dia}) . Factors $C_n\Lb \gamma\Rb\Phi_n\Lb e^{- \gamma},\gamma\Rb$ are the residues (impact factors) of the Green's function. Each term is intimately related to the inclusive production of one, two and so on particles\cite{MU70}, and, therefore, has a direct physics meaning. The series of \eq{SA1} is a typical example of the asymptotic series, sum of which is the analytical function that has the same expansion. In our paper \cite{utmp} we proposed the way of summing this asymptotic series.

 ~

\subsection{Multiplicity distribution for the 'dresssed' Pomeron}

\begin{figure}[ht]
    \centering
  \leavevmode
      \includegraphics[width=8cm]{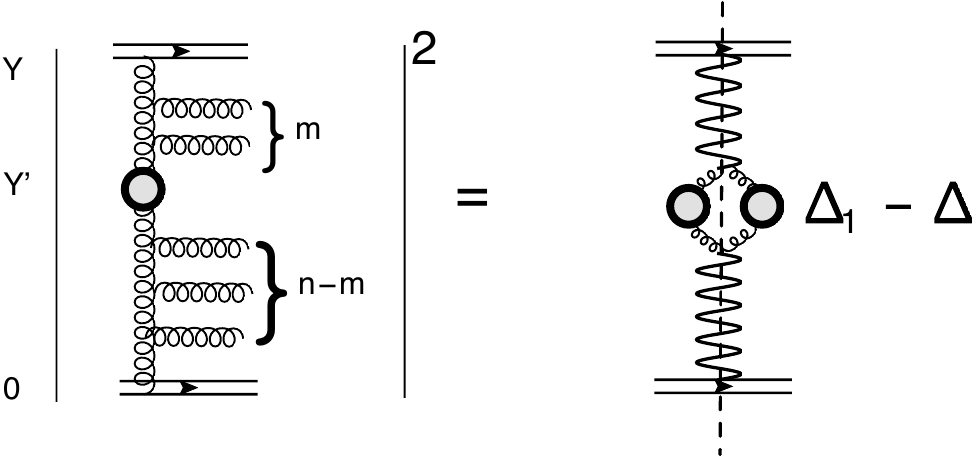}  
      \caption{ The multiplicity distribution for the first enhanced diagram in the 'dressed' Pomeron. The wavy line denotes the cut Pomeron. Helix lines describe the partons.  $\Delta_1 = e^\gamma - 1$ and $\Delta = \gamma$ is the intercept of the 'bare' Pomeron.}
\label{diaprpom}
   \end{figure}

In section 3.2 we assumed that the 'dressed' Pomeron has the same Poisson
distribution of the produced dipoles as the 'bare' Pomeron. \fig{diaprpom} illustrates this point of view considering the first enhanced diagram for the 'dressed' Pomeron. 

  The $\sigma_n$ for this diagram is equal to
  \beq \label{PC13}
  \sigma_n =\intl^Y_0 d  Y'\gamma \Lb \Delta_1 - \Delta\Rb \sum_{m=1}^{n} \frac{ \Lb \Delta \Lb Y - Y'\Rb\Rb^m}{m!} e^{ - \Delta\Lb Y - Y'\Rb} \frac{ \Lb \Delta \Lb Y - Y'\Rb\Rb^{n - m}}{\Lb n - m\Rb!} e^{ - \Delta\,Y'} = 
  \Lb \Delta_1 - \Delta\Rb Y  \frac{ \Lb \Delta Y\Rb^n}{n!} e^{ - \Delta\ Y} 
  \eeq
  Sum of all enhanced diagrams leads to
  \beq \label{PC14}
  \frac{\sigma^{\pom}_n}{ \sigma^{\pom}_{in} }\,=\, \frac{ \Lb \Delta Y\Rb^n}{n!} e^{ - \Delta\ Y}\eeq
  \eq{PC14} is the Poisson distribution, the same as for the' bare' Pomeron in \eq{I2}.

\end{document}